\documentclass[%
 reprint,
preprintnumbers,
nofootinbib,
prd,
]{revtex4-2}

\usepackage{graphicx}
\usepackage{dcolumn}
\usepackage{natbib}
\usepackage{bm}
\usepackage{hyperref}
\usepackage{feynmf}
\usepackage{graphicx}
\usepackage{enumitem}
\usepackage{latexsym}
\usepackage{amsfonts}
\usepackage{amssymb}
\usepackage{mathrsfs}
\usepackage{color}
\usepackage{amsmath}
\usepackage{slashed}
\usepackage{dcolumn}
\usepackage{verbatim}
\usepackage{comment}
\usepackage{float}
\usepackage{multirow}
\usepackage[normalem]{ulem}
\usepackage[dvipsnames]{xcolor}
\usepackage{mathtools}
\usepackage{appendix}

\definecolor{sfugreen}{HTML}{1D8038}

\begin{document}

\preprint{APS/123-QED}

\title{Searching for dark matter signals with high energy astrophysical neutrinos in IceCube
} 

\author{Khushboo Dixit$^{1}$}
\email{kdixit@uj.ac.za}

\author{Gopolang Mohlabeng$^{2,3}$}
\email{gmohlabe@sfu.ca}

\author{Soebur Razzaque$^{1,4,5}$}
\email{srazzaque@uj.ac.za}


\affiliation{$^{1}$Centre for Astro-Particle Physics (CAPP) and Department of Physics,
University of Johannesburg, PO Box 524, Auckland Park 2006, South Africa}

\affiliation{$^{2}$Department of Physics, Simon Fraser University,
Burnaby, BC, V5A 1S6, Canada}

\affiliation{$^{3}$TRIUMF, 4004 Westbrook Mall,
Vancouver, BC, V6T 2A3, Canada}

\affiliation{$^{4}$Department of Physics, The George Washington University,
Washington, DC 20052, USA}

\affiliation{$^{5}$National Institute for Theoretical and Computational Sciences (NITheCS),
Private Bag X1, Matieland, South Africa}


\date{\today}

\begin{abstract}
High-energy neutrinos provide a potentially powerful and  distinctive probe for dark matter (DM) - neutrino interactions, particularly in environments with enhanced DM densities, such as the DM spikes predicted to form around supermassive black holes (SMBHs) at the center of active galactic nuclei (AGN). Recent results by the IceCube Neutrino Observatory, which reported four significant AGNs, namely TXS 0506+056, NGC 1068, PKS 1424+240, and NGC 4151 as candidate neutrino sources, provide a valuable opportunity to search for signatures of these interactions. In this study, we use IceCube data to derive the most stringent constraints to date on both the energy-dependent and energy-independent DM-neutrino scattering cross-sections. We perform a statistical analysis using data from individual sources as well as a combined (stacked) analysis of all four sources.
Our strongest limits arise from the stacking analysis, yielding an upper bound of $\sigma_{0} \lesssim 8\times 10^{-39}$ cm$^2$ for an energy-independent cross-section and $\sigma_{0} \lesssim 10^{-39}$ cm$^2$ for a linearly energy-dependent cross-section, both at 90\% confidence level, particularly in scenarios involving the adiabatic growth of black holes. 

\end{abstract}

\maketitle

\section{\label{sec:Introduction} Introduction} 

The existence of dark matter (DM) is well-established through its gravitational interactions with standard model (SM) matter in astrophysical and cosmological systems \cite{Bertone:2016nfn, cirelli2024dark}. Yet, its fundamental nature remains unknown. Discovery of a clear signal necessitates a multi-pronged approach that integrates novel theoretical techniques with a broad experimental search program. While both neutrinos and DM may interact only feebly with the rest of the SM, it is reasonable to explore the possibility that they interact strongly with each other. 
Among the many probes, high-energy astrophysical neutrinos present a remarkable opportunity to investigate possible strong interactions between these two sectors, across vast astrophysical distances \cite{Arguelles:2017atb, Choi:2019ixb, Hooper:2021rjc, PhysRevLett.130.091402, Cline_2023,ferrer2023new}. A fraction of these high-energy neutrinos are created in the extreme and dense environments of AGN, which harbor supermassive black holes (SMBHs), and where conditions are ripe for acceleration and interactions of cosmic-ray protons and nuclei \cite{keivani2018multimessenger, Reimer:2018vvw, Rodrigues:2018tku, Cerruti:2018tmc, Inoue_2019, Das:2021cdf, Das:2022nyp}.
As these neutrinos propagate through space, interactions with DM, in regions of high DM density, may lead to observable scattering or attenuation effects, particularly in the intriguing and predicted DM density spikes surrounding SMBHs \cite{PhysRevLett.130.091402,Cline_2023}.

Recently, the IceCube Neutrino Observatory has made groundbreaking strides in neutrino astronomy by indicating several significant extragalactic point sources. IceCube released its results from a recent time-integrated analysis of 10-year data reporting aggregated neutrino emission observation from a list of 110 gamma-ray sources \cite{IceCube:2021slf}. The increase in neutrino events has been observed notably from three main candidate sources, including TXS~0506+056 \cite{IceCube:2018dnn, IceCube:2018cha}, NGC~1068 \cite{icecube2022evidence}, PKS~1424+240 \cite{icecube2022evidence}. Later on, one more significant candidate source, NGC~4151 \cite{abbasi2025search}, was reported in \cite{IceCube:2024ayt}. These findings represent a major advancement, allowing for further direct tests of DM-neutrino scattering.
In this study, we leverage the results presented in \cite{dixit2024searching} from publicly available IceCube data \cite{icecube2022evidence} to derive constraints on the DM-neutrino scattering cross-section, considering both energy-independent and energy-dependent interaction scenarios.
To this end, we analyze individual sources and perform a stacking analysis that integrates their results to obtain comprehensive bounds on DM-neutrino scattering.

Previous studies have used astrophysical neutrino data from TXS~0506+056 \cite{PhysRevLett.130.091402,ferrer2023new} and NGC~1068 \cite{Cline_2023} to constrain the DM-neutrino scattering cross-section. In these analyses, the 90\% confidence level (CL) upper limits on the cross-section were derived by adopting the corresponding 90\%~CL lower limits on the number of neutrino events observed by IceCube for each source. 
In this work, we instead perform a full statistical treatment to derive constraints on the DM–neutrino cross-section, both for individual sources and through a stacked analysis of all four sources.

The rest of this article is structured as follows:
In section \ref{sec:AstroNeutrinos} we report the astrophysical neutrino data in terms of the number of events observed from each source. 
We then discuss the spike density models predicting the DM distribution around SMBHs in section \ref{sec:DarkMatter}. In section \ref{sec:DmNeutrino}, we discuss the neutrino flux attenuation due to DM-neutrino scattering as the neutrinos propagate from AGN to the earth. 
In section \ref{sec:DataAnalysis}, we provide the details of the statistical analysis. We then present our model independent results in section \ref{sec:DataAnalysis}, followed by a model interpretation in section \ref{sec:model}.
Finally, we conclude in section \ref{sec:Conclusions}.

\section{\label{sec:AstroNeutrinos} High Energy Neutrinos from Astrophysical Sources} 
IceCube has been searching for point-like sources of astrophysical neutrinos, such as AGN and blazars, using its PStrack events that are induced by astrophysical muon neutrinos ($\nu_\mu$) and anti-neutrinos ($\bar{\nu}_\mu$). Among the most significant IceCube source candidates in the 10-year IC86 point-source analyses are NGC 1068, TXS 0506+056, and PKS 1424+240, each reported at a source-level local significance above 3.3$\sigma$ \cite{IceCube:2021slf}. For NGC 1068 specifically, IceCube later reported an individual source global significance of 4.2$\sigma$ \cite{icecube2022evidence}. In the same work, IceCube also quoted a catalog-level binomial-test significance for the combined set {NGC 1068, PKS 1424+240, TXS 0506+056} of about 3.7$\sigma$ local (3.4$\sigma$ global). IceCube further reported evidence for neutrino emission from the Seyfert galaxy NGC 4151 at a significance of 2.9$\sigma$ \cite{IceCube:2024ayt}. 
These results have been based on the assumption that neutrino fluxes from these sources follow an unbroken power-law (UP) of the form
\begin{equation}\label{eq:nuflux}
    \Phi_{\nu_\mu+\bar{\nu}_\mu} = \Phi = \Phi_{ref}\bigg(\frac{E_{\nu}}{E_0}\bigg)^{-\Gamma},
\end{equation}
where $\Phi_{ref}$ is the normalization flux, $\Gamma$ is the spectral index, and $E_{\nu}$ and $E_0$ are the energy of neutrinos and the reference energy, respectively. 
In a previous work \cite{dixit2024searching}, we conducted a likelihood analysis of the IceCube public data provided in \cite{icecube2022evidence}, where, following the UP in Eq. (\ref{eq:nuflux}), we presented the observed number of events $n_s$ and the spectral indices for all four sources. The results are also summarized in Table \ref{tab:n_s}.

\begin{table}[htbp]
\caption{\label{tab:n_s}%
Best fit $n_s$ and $\Gamma$ along with 1$\sigma$ errors observed for the neutrino energy range of 0.1 TeV - 1 PeV for each source mentioned in this work \cite{dixit2024searching}.
}
\begin{ruledtabular}
\begin{tabular}{lcc}
\textrm{Source} &
\textrm{$n_s \pm 1\sigma$} &
\textrm{$\Gamma \pm 1\sigma$} \\
\colrule

NGC 1068        & $75^{+16}_{-15}$ & $2.75^{+0.2}_{-0.2}$ \\
TXS 0506+056    & $28^{+12}_{-12}$ & $2.2^{+0.3}_{-0.2}$ \\
PKS 1424+240    & $42^{+16}_{-14}$ & $3.0^{+0.5}_{-0.4}$ \\
NGC 4151        & $30^{+13}_{-10}$ & $2.4^{+0.4}_{-0.3}$ \\
\end{tabular}
\end{ruledtabular}
\end{table}

\section{\label{sec:DarkMatter}DM Distribution Around SMBHs}

In this section, we discuss the possibility of an enhanced DM distribution around SMBHs at the center of active galaxies. In early theoretical and empirical models, 
the DM density in the region close to SMBH was approximated as cuspy profile of the form \cite{Navarro:1995iw,Kravtsov:1997vm}
\begin{equation}\label{eq:cusp}
    \rho \simeq \rho_s \left(\frac{r}{r_s}\right)^{-\gamma},
\end{equation} 
within the small range of radii $r$. Here, $\rho_s$ and $r_s$ are the characteristic density and scale radius, respectively and are defined as \cite{Navarro:1995iw,Bryan:1997dn,WMAP:2006bqn}
\begin{equation}
    \rho_s = \frac{\Delta}{3} \frac{\eta^3}{\ln (1+\eta) - \frac{\eta}{1+\eta}}\rho_c; ~~~~r_s = 8.8 \bigg(\frac{M_{vir}}{10^{11}M_{\odot}}\bigg)^{0.46}~ {\rm kpc}.
\end{equation}
The quantity $\rho_c = 9\times 10^{-27}$~kg/m$^3$ is the critical density of the Universe, $\Delta$ is the virial over density $\approx300$ \cite{Bryan:1997dn}, $M_{vir}$ is the virial mass, and the concentration parameter $\eta$ of the NFW halo is parameterized as
\begin{equation*}
    \eta \simeq 13.6 \bigg({\frac{M_{vir}}{10^{11}M_{\odot}}}\bigg)^{-0.13}.
\end{equation*}
However, it has been reported that the possibility of adiabatic accretion of DM gives this profile a spiky nature that can be described as follows \cite{Gondolo:1999ef}
\begin{equation}
    \rho_{\chi} = \rho_R\bigg(1-\frac{4R_s}{r}\bigg)^3\bigg(\frac{R_{sp}}{r}\bigg)^\alpha,
    \label{eq:spike_dens}
\end{equation}
where, $R_s = 2GM_{BH}/c^2$ is the Schwarzschild radius, $R_{sp}$ is the DM spike radius, and $\rho_R \approx \rho(R_{sp})$ (using Eq. \ref{eq:cusp}) represents the initial cuspy DM density profile at $r=R_{sp}$. 
The spectral index $\alpha$ is related to the parameter $\gamma$ of the initial cuspy profile as $\alpha = (9-2\gamma)/(4-\gamma)$. 
On the other hand, a cuspy DM density profile (the NFW-profile) \cite{Navarro:1995iw} was established across a wide range of halo masses, ranging from dwarf galaxies to rich clusters, within the Cold Dark Matter (CDM) cosmological framework. This profile scales with halo radius (r) as 
\begin{equation}
    \rho_{NFW}(r) = \rho_s\bigg(\frac{r}{r_s}\bigg)^{-\gamma}\bigg(1+\bigg(\frac{r}{r_s}\bigg)^\xi\bigg)^{-(\beta - \gamma)/\xi},
\end{equation}
where $(\xi, \beta, \gamma) = (1,3,1)$ \cite{Zhao:1995cp,Stahl:2024jzk}.
Here, $\gamma$ is the inner slope that controls the cusp/core behavior near the center, $\beta$ is the outer slope that determines the falloff at the radius $r$, and $\xi$ defines the sharpness of the transition between the inner and outer regimes. 
For the value of $\gamma = 1$, as found in citation for the NFW profile, the resulting spike index is $\alpha = 7/3$, which determines the spike density profile in Eq.~\ref{eq:spike_dens}.

Another possibility is gravitational scattering between the DM and dense stellar components, which is likely to be present in the inner region of the galaxy. This scenario relaxes the spike profile by lowering the spectral index to $\alpha = 3/2$ independent of any value of $\gamma$ \cite{PhysRevLett.92.201304,PhysRevLett.93.061302,PhysRevD.75.043517,PhysRevD.106.043018}. 

A precise incorporation of these two possibilities \cite{Gondolo:1999ef,PhysRevLett.93.061302,Vasiliev:2008uz} can be represented as
\begin{equation}
\rho_\chi^{3/2}(r) \simeq
\begin{cases}
\rho_{N}\!\left(1 - \dfrac{4 R_{S}}{r}\right)^{3}\!\left(\dfrac{r_{h}}{r}\right)^{3/2}, & r_{i} \le r \le r_{h},\\[6pt]
\rho^\prime_{N}\!\left(\dfrac{R_{\mathrm{sp}}}{r}\right)^{7/3}, & r \ge r_{h}\,,
\end{cases}
\label{eq:3.3}
\end{equation} 
where $r_i = 4R_s$ is the inner radius of the spike, and $r_h = G M_{BH}/\sigma^{\ast 2}$ is the influence radius of SMBH with $\sigma^{\ast}$ being the stellar velocity dispersion. The effects of DM scattering with stellar components are significant only within this radius \cite{Gondolo:1999ef}. 
$\rho_N$ and $\rho_N^\prime$ are normalization constants. Their evaluation is discussed later in this section. For $r > r_h$, the spike profile converges to 
\begin{equation}
    \rho_\chi^{7/3}(r) \simeq  \rho_{N}\!\left(1 - \dfrac{4 R_{S}}{r}\right)^{3}\!\left(\dfrac{R_{sp}}{r}\right)^{7/3}, ~~~~~r \geq r_{i}. 
\end{equation}

\begin{table*}[htbp]
\caption{\label{tab:parameters}List of parameters used in this work associated with each source. We have also reported the $\Sigma_\chi$ values, in units of GeV/cm$^2$, for each benchmark model from all sources for $m_\chi = 1$ GeV and $r=10^4$ pc in this table.
}
\begin{ruledtabular}
\begin{tabular}{lcccc}
 Parameters & NGC 1068 & NGC 4151 & TXS 0506+056
& PKS 1424+240\\ \hline 
 $M_{BH}$ ($M_\odot$) &$1.0\times 10^{7}$\cite{Greenhill:1996te} &$2.0\times 10^{7}$\cite{Bentz_2022} &$3.09\times 10^{8}$\cite{Padovani:2019xcv} & $1.0\times 10^{9}$\cite{Cerruti:2017mnz} \\
 $t_{BH}$(yr)&$10^9$
 &$10^9$ & $10^9$ & $10^9$\\
 $R_s$(pc) & $9.5\times 10^{-7}$ & $1.9\times 10^{-6}$ & $2.95\times 10^{-5}$ & $9.5\times 10^{-5}$ \\
 $r_h$(pc)  & $6.5\times 10^{5} R_s$ & $33.4\times 10^{5} R_s$ & $10^{5} R_s$ & $10^{5} R_s$ \\
 $R_{em}$(pc) & $30 R_s$ & $30 R_s$ & $2.2\times 10^{3} R_s$ & $6.7\times 10^{2} R_s$ \\

 $\Sigma_{\chi{\rm -BM1}}$ & $5.77\times 10^{31}$ & $2.88\times 10^{31}$ & $2.69\times 10^{28}$ & $3.95\times 10^{28}$ \\
 $\Sigma_{\chi{\rm -BM2}}$  & $2.45\times 10^{28}$ & $2.71\times 10^{28}$ & $2.27\times 10^{28}$ & $3.15\times 10^{28}$ \\
 $\Sigma_{\chi{\rm -BM3}}$ & $9.45\times 10^{26}$ & $1.04\times 10^{27}$ & $2.42\times 10^{27}$ & $2.91\times 10^{27}$ \\
 $\Sigma_{\chi{\rm -BM1^\prime}}$ & $9.28\times 10^{28}$ & $4.64\times 10^{28}$ & $6.04\times 10^{27}$ & $3.49\times 10^{27}$ \\
 $\Sigma_{\chi{\rm -BM2^\prime}}$  & $9.91\times 10^{27}$ & $7.95\times 10^{27}$ & $5.96\times 10^{27}$ & $3.47\times 10^{27}$ \\
 $\Sigma_{\chi{\rm -BM3^\prime}}$ & $1.11\times 10^{27}$ & $1.0\times 10^{27}$ & $2.39\times 10^{27}$ & $1.78\times 10^{27}$ \\
\end{tabular}
\end{ruledtabular}
\end{table*}
At sufficiently high DM densities, DM self-annihilations become significant. 
The process leads to a flattening of the DM spike towards the innermost region, in which the density reaches a saturation (maximum) value, $\rho_{sat} = m_\chi /(\langle \sigma_a v\rangle t_{BH})$.
Here, $m_{\chi}$ is the DM mass, $\langle \sigma_a v\rangle$ is the thermally averaged annihilation cross section of DM, and $t_{BH}$ is the age of the SMBH.
Taking these effects into account, the resulting DM spike density profile is expressed as \cite{Gondolo:1999ef}
\begin{equation}
\rho_{\chi}(r) =
\begin{cases}
0, & r \le r_i,\\[6pt]
\dfrac{\rho_\chi^{\alpha}(r)\,\rho_{sat}}{\rho_\chi^{\alpha}(r) + \rho_{sat}}, & r_i \le r \le R_{\mathrm{sp}},\\[10pt]
\dfrac{\rho_{\mathrm{NFW}}(r)\,\rho_{sat}}{\rho_{\mathrm{NFW}}(r) + \rho_{sat}}, & r \ge R_{\mathrm{sp}}\,,
\end{cases}
\label{eq:3.6}
\end{equation}
where $\alpha=7/3,3/2$. In our analysis, we consider the benchmark models as given below \cite{granelli2022blazar}
\begin{itemize}
    \item BM1, BM2, BM3: choose $\alpha = 7/3$ with $\langle \sigma_a v\rangle = (0, 10^{-28}, 3\times 10^{-26})$ cm$^3$~sec$^{-1}$, respectively.
    \item BM1$^\prime$, BM2$^\prime$, BM3$^\prime$: same annihilation cross-sections but with $\alpha = 3/2$.
\end{itemize}
The choice $\langle \sigma_a v \rangle = 0$ is applicable to scenarios such as asymmetric DM. In contrast, the value $\langle \sigma_a v \rangle = 3\times 10^{-26} \text{ cm}^3\text{ sec}^{-1}$ corresponds to the canonical expectation for thermal relic DM.

\subsection{Profile parameterization}

\begin{figure*}[htbp]
\includegraphics[width=0.44\linewidth]{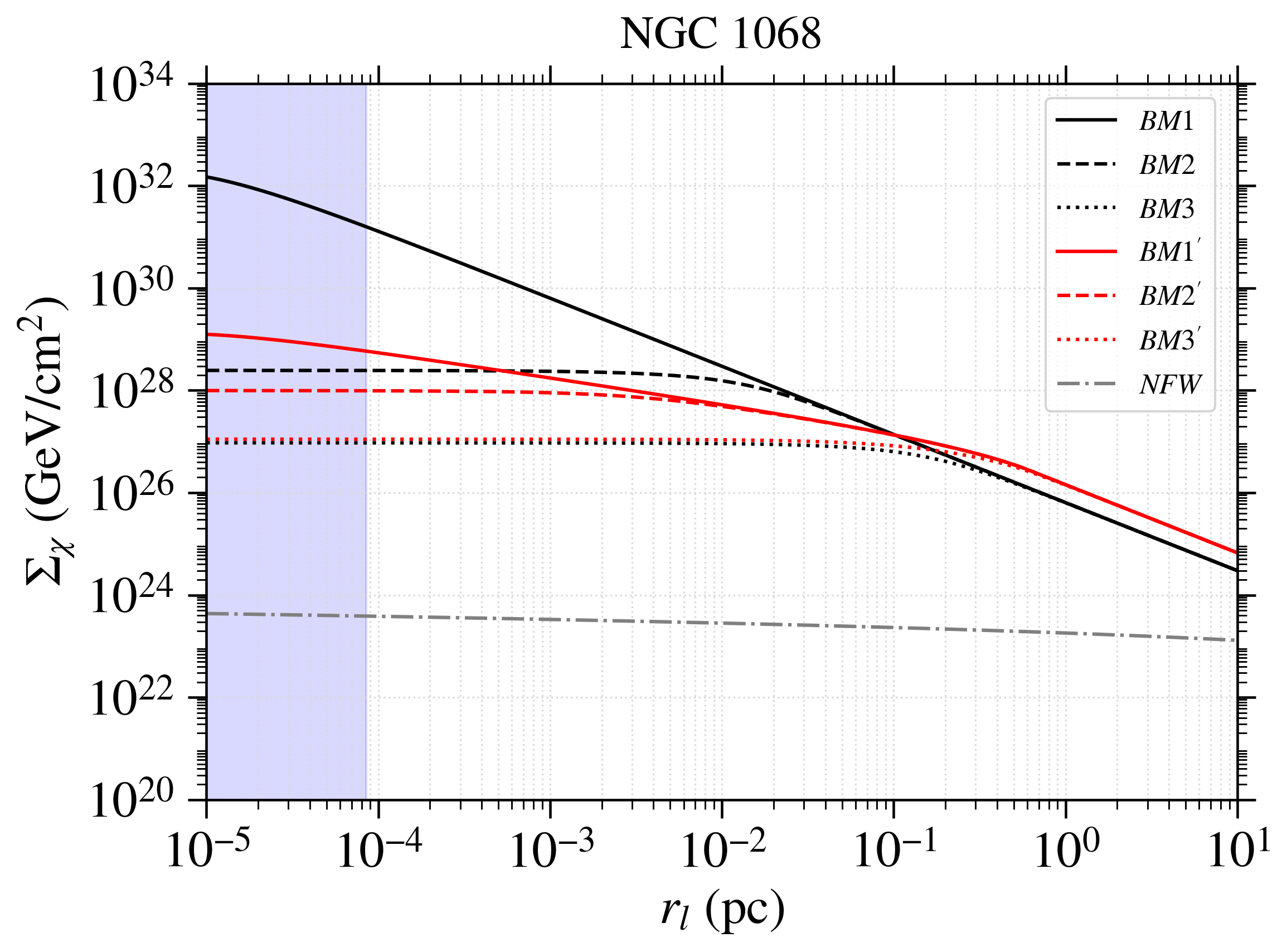}
\includegraphics[width=0.44\linewidth]{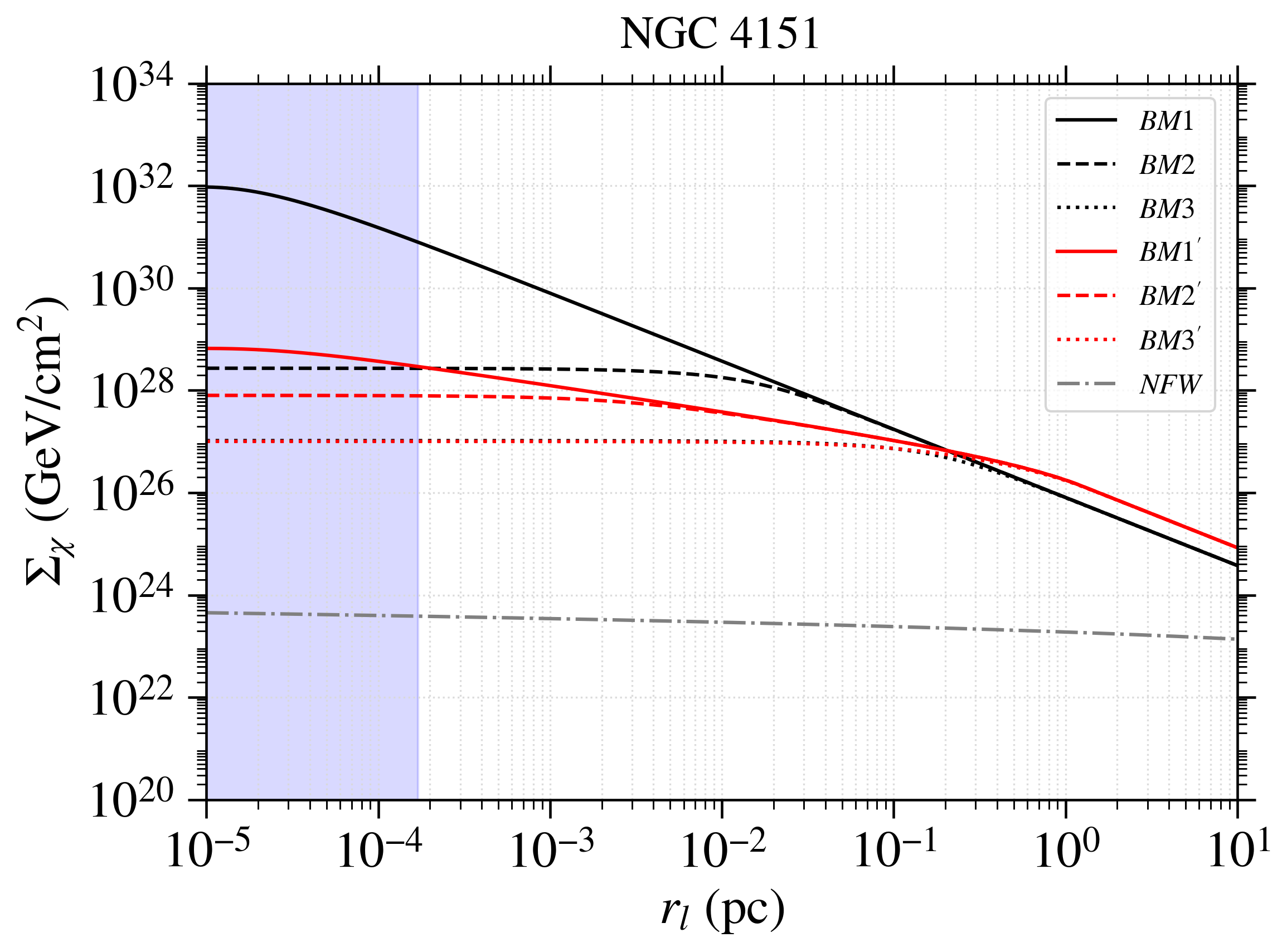}
\includegraphics[width=0.44\linewidth]{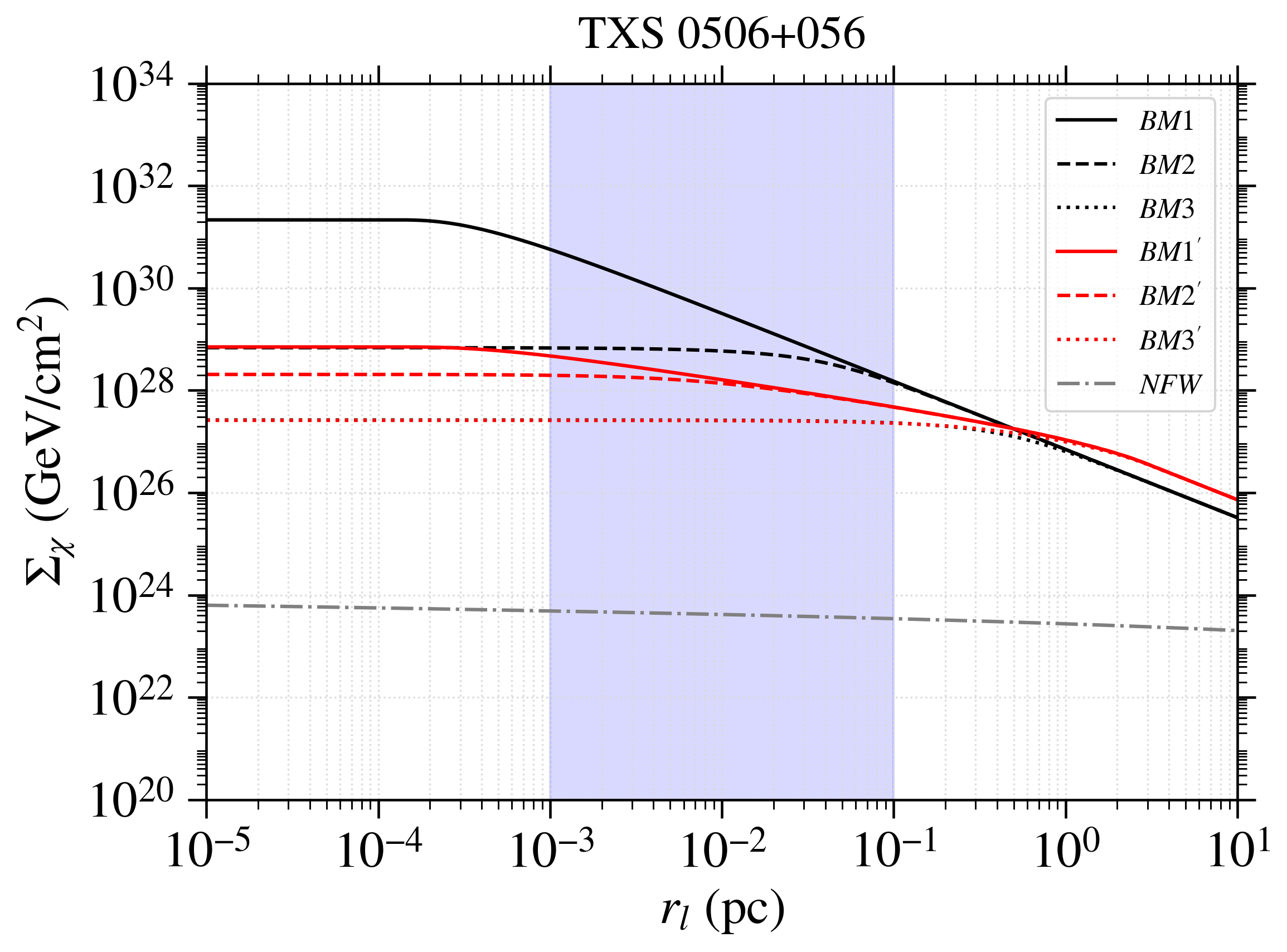}
\includegraphics[width=0.44\linewidth]{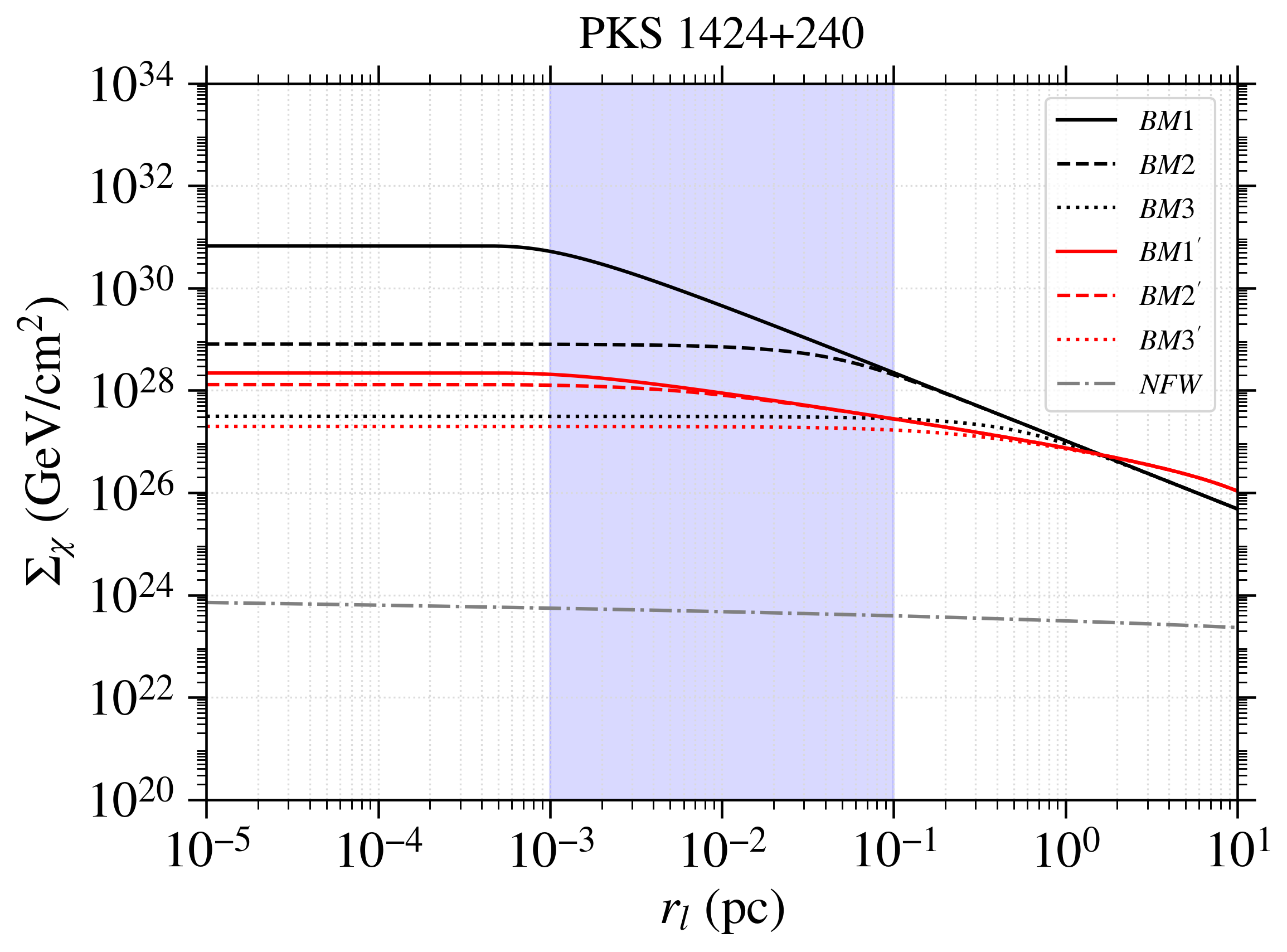}
\caption{DM column density $\Sigma$ Vs. $r_l$ (the lower limit to the integration given in Eq. (\ref{eq:Sigma_chi})) for each source. We considered the upper limit to the same integration $r=10^4$ pc and $m_\chi = 1$ GeV to obtain these plots. The shaded regions represent the possible range of $R_{\rm em}$ for each source. 
} 
\label{fig:SigmaVsrl}
\end{figure*}

The values of the parameters we have used in this work are provided in Table \ref{tab:parameters}, for each source. For NGC 1068, where low-energy neutrinos are expected to originate primarily from hadronic interactions in the inner dusty torus or corona, we adopt an emission radius of $R_{em} = 30R_s$, following \cite{Cline_2023,Murase:2019vdl}. 
In the absence of direct observational constraints, a physically motivated range for these Seyfert systems spans a few tens of Schwarzschild radii \cite{Inoue:2019yfs,Inoue_2019,Murase:2019vdl,Inoue:2018kbv,Gallimore:2004wk,Murase:2022dog}. 
Owing to their comparable accretion structure and central engine geometry, we apply the same choice of $R_{em}$ to NGC 4151. The same argument also motivates using an identical $R_{em}$ for the blazars TXS 0506+056 \cite{PhysRevLett.130.091402} and PKS 1424+240, which are both gamma-ray blazars with similar black hole masses.  
Throughout our analysis, we assume a black-hole age of $10^9$ yr, a commonly adopted and well-supported timescale based on \cite{10.1093/mnras/staa3363}. The parameter $r_h$ has been taken from \cite{Greenhill:1996te,Hure:2002nu,Lodato:2002cv,Woo:2002un,Panessa:2006sg} for NGC~1068, while for NGC~4151, we use the stellar dispersion velocity observed to be $\sigma^{\ast 2} = 116 \pm 3$ km/sec in \cite{2014ApJ...791...37O} to calculate $r_h$. In the absence of any information on $r_h$ or $\sigma^\ast$ for TXS~0506+056 and PKS~1424+240, we have assumed $r_h = 10^5R_s$, the typical radius relevant for BH mass estimations \cite{PhysRevD.82.083514}. 
To calculate the normalizations $\rho_N$ and $\rho_N^\prime$ we need the mass of the BH, which is given as
\begin{equation}
    M_{BH} \approx 4\pi \int_{r_i}^{r_h} dr r^2 \rho_\chi(r).
\end{equation}
Here, $r_h$ is the influence radius of the BH, while the DM density $\rho_\chi(r) \simeq \rho_\chi^\alpha$ ($\alpha = 7/3,3/2$) for $r~\gg~r_i$, given by Eqs. 6 and 7.
To calculate $\rho_N$ we use the relation $\mathcal{M}\approx\rho_N r_h^{3/2}$ or $\mathcal{M}\approx\rho_N R_{sp}^{7/3}$. Under the condition $\rho_\chi(r) \simeq \rho_\chi^\alpha$ one can write
\begin{equation}
\mathcal{M}\simeq \frac{M_{BH}}{4\pi \big(f_\alpha(r_h) - f_\alpha(r_i)\big)},
\end{equation}
where,
\begin{equation}
    f_\alpha(r) \equiv r^{-\alpha}\bigg[\frac{r^3}{3-\alpha}+\frac{12 R_s r^2}{\alpha-2}-\frac{48 R_s^2 r}{\alpha-1}+\frac{64 R_s^3}{\alpha}\bigg].
\end{equation}
This expression follows from integrating the binomial expansion of the factor
\[
\rho(r)\propto r^{2-\alpha}\left(1-\frac{4R_S}{r}\right)^3,
\]
where the term $\left(1-4R_S/r\right)^3$ encodes the relativistic suppression of phase-space orbits near the BH. The leading contribution, proportional to $r^{2-\alpha}$, corresponds to the standard Newtonian adiabatic spike term, while the remaining terms scaling as $r^{1-\alpha}$, $r^{-\alpha}$ and $r^{-1-\alpha}$ arise from the linear, quadratic, and cubic components of the expansion of $(1 - 4R_S/r)^3$, and therefore represent the first, second, and third order general relativistic (GR) corrections to the spike structure respectively \cite{Gondolo:1999ef, Sadeghian:2013laa}.

However, we do not consider the relativistic corrections to the spike as proposed in \cite{Sadeghian:2013laa} because they are negligible at distances $\gtrsim 20$ pc \cite{Cline_2023}. Indeed for all sources examined here, the value of \(R_{em}\) exceeds this limit. 
$\rho_N^\prime$ is calculated using the relation $\rho_N^\prime \simeq \rho_N (r_h/R_{sp})^{7/3}$ ensuring that the profile $\rho_\chi^{3/2}$ matches at the intersection point $r_h$. In addition, the spike radius can be calculated using the condition that the DM halo should continue following the NFW profile outside of the spike radius by imposing the condition $\rho_N^\prime, \rho_N \simeq \rho_s (R_{sp}/r_s)^{-\gamma}$. Hence, we define the spike radii as $R_{sp}^{3/2} = (\mathcal{M}/\rho_s r_s)^{3/4}~r_h^{5/8}$ and $R_{sp}^{7/3} = (\mathcal{M}/\rho_s r_s)^{3/4}$.

The parameters $r_s$ and $\rho_s$ defining the NFW profile, are calculated based on the virial mass, which is taken as the DM halo mass $M_{DM}$ and is calculated using the relation of $M_{DM}$ with the SMBH mass as
\begin{equation}
    M_{BH} \sim 7\times 10^7 M_{\odot}\bigg(\frac{M_{DM}}{10^{12}M_{\odot}}\bigg)^{4/3},
\end{equation}
This expression has been predicted using N-body simulations in a $\Lambda$CDM universe \cite{DiMatteo:2003zx} and is consistent with results found in \cite{Ferrarese:2002ct, Baes:2003rt}. 

The emission point of neutrinos $R_{em}$ near the SMBH is taken as $30\,R_s$ for NGC 1068 and NGC 4151 because of lower-energy ($< 10$~TeV) neutrinos detected from these sources. This requires relatively lower-energy protons to interact with a higher-energy radiation field or dense gas, both of which are found relatively closer to the SMBH, (see e.g., \cite{Inoue:2019yfs}). While for TXS 0506+056, neutrinos having higher energy (290 TeV) are supposedly produced at a comparatively larger distance from the SMBH, where the radiation field is at relatively lower energy, thus interacting with high-energy protons, which is consistent with the lepto-hadronic model of high energy neutrino production in blazars \cite{Das:2022nyp}.

\subsection{DM column density}\label{columnden}
Combining the ingredients described in the previous sections, we evaluate the DM column density, which quantifies the probability for neutrinos to scatter off DM as they propagate towards the Earth. 
The column density $\Sigma_\chi$ is obtained by integrating $\rho_\chi$ over an estimated distance traveled by neutrinos, expressed as 
\begin{equation}\label{eq:Sigma_chi}
    \Sigma_\chi(r) = \int_{r_l}^{r} dr^\prime \rho_\chi(r^\prime) \,.
\end{equation}
Here, $r_l = R_{em}$ with $r=10^4$ pc, the upper limit of the integral given in Eq. (\ref{eq:Sigma_chi}). Above this upper limit, we obtain an approximately constant $\Sigma_\chi$ for each source. It is shown in Fig.~\ref{fig:SigmaVsr} of Appendix \ref{sec:column}, where we plot $\Sigma_\chi(r)$ vs.\ $r$. At $r>10^4$ pc, there is a negligible change in $\Sigma_\chi$ for each benchmark model. For the NFW profile, $\Sigma_\chi$ ranges between (3.5.-4.1)$\times 10^{23}$ GeV/cm$^2$ for $r=10^4$ pc.  However, a very small increase, of the order of $10^{-2}$ (thus negligible), can be seen in the NFW profile for $r=10^6$ pc. Figure \ref{fig:SigmaVsrl} shows the variation of column density $\Sigma_{\chi}$ vs.\ $r_l$ for each source. 

In this analysis, a key consideration is the uncertainty associated with the neutrino emission point $R_{\rm em}$. We selected specific values for $R_{\rm em}$ (see Table \ref{tab:parameters}) based on previous analyses \cite{PhysRevLett.130.091402,Cline_2023}. 
In Fig. \ref{fig:SigmaVsrl}, we observe how $\Sigma_\chi$ varies with different values of $R_{\rm em}$. The shaded blue region represents the allowed range of $R_{\rm em}$ for each source. For instance, for NGC~1068 and NGC~4151, the range is $10R_s \lesssim R_{\rm em} \lesssim 90R_s$ \cite{Murase:2022dog}. In contrast, for TXS~0506+056 and PKS~1424+240, we considered $R_{\rm em}$ in the range of $(10^{-3}-10^{-1})$ pc as indicated in Ref.~\cite{Padovani:2019xcv}. The constraints for NGC~1068 and NGC~4151 can vary significantly within the specified range of $R_{\rm em}$ for the BM1 profile, potentially by an order of magnitude. A minor variation is observed in $\Sigma_\chi$ for BM1$^\prime$; however, $\Sigma_\chi$ for other profiles remain almost constant within the considered $R_{\rm em}$ range. For TXS~0506+056 and PKS~1424+240, the BM1 spike may vary within nearly three orders of magnitude, while the BM1$^\prime$ shows a decrease of about one order of magnitude. Additionally, slight variations are noted in BM2 and BM2$^\prime$, whereas the spikes for BM3 and BM3$^\prime$ are almost constant. Since the bound on $\sigma_0$ is inversely proportional to the column density $\Sigma_\chi$, a decrease in $\Sigma_\chi$ will relax the bound on $\sigma_0$ by the same order of magnitude. 

\section{\label{sec:DmNeutrino}Neutrino flux attenuation due to interaction with DM}

The neutrino flux attenuation due to scattering with DM is evaluated using the cascade equation \cite{Arguelles:2017atb,Vincent:2017svp}
\begin{equation}
\frac{d\Phi}{d\tau}(E_\nu)
= -\,\sigma_{\nu\chi}\,\Phi(E_\nu)
+ \int_{E_\nu}^{\infty} dE^\prime_\nu\,
\frac{d\sigma_{\nu\chi}}{dE_\nu}(E^\prime_\nu \!\to\! E_\nu)\,
\Phi(E^\prime_\nu)\,,
\label{eq:cascade}
\end{equation}
where, $\tau = \Sigma_\chi(r)/m_\chi$ is the accumulated column density over the DM mass, $E_\nu$ and $E_\nu^\prime$ are the outgoing and incoming neutrino energies, respectively. The first term in this equation represents the attenuation of the neutrino flux, while the second term represents the redistribution of high energy neutrinos in the lower energy bins, due to scattering with DM.

\begin{figure*}[htbp]
\includegraphics[width=0.46\linewidth]{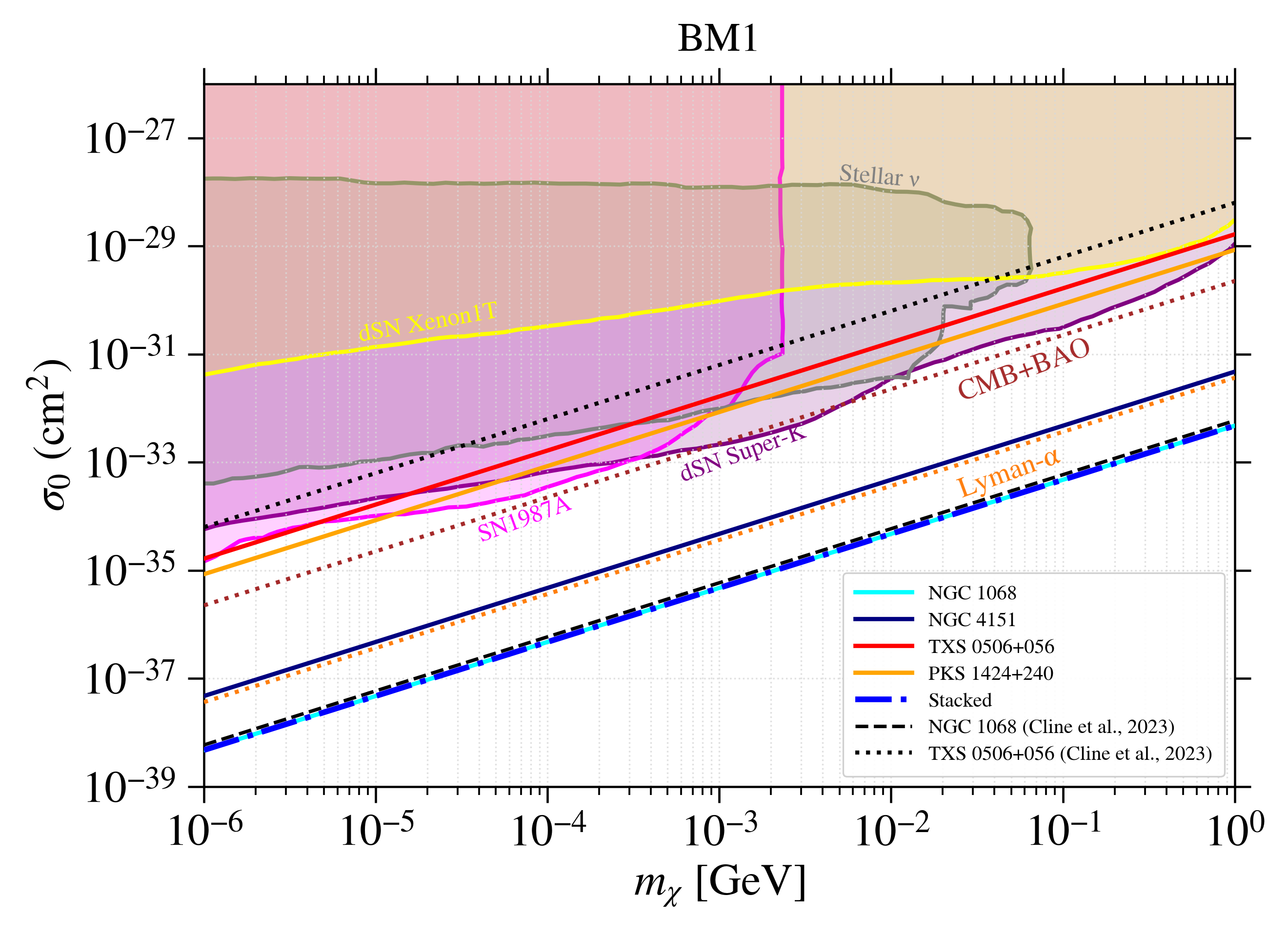}
\includegraphics[width=0.46\linewidth]{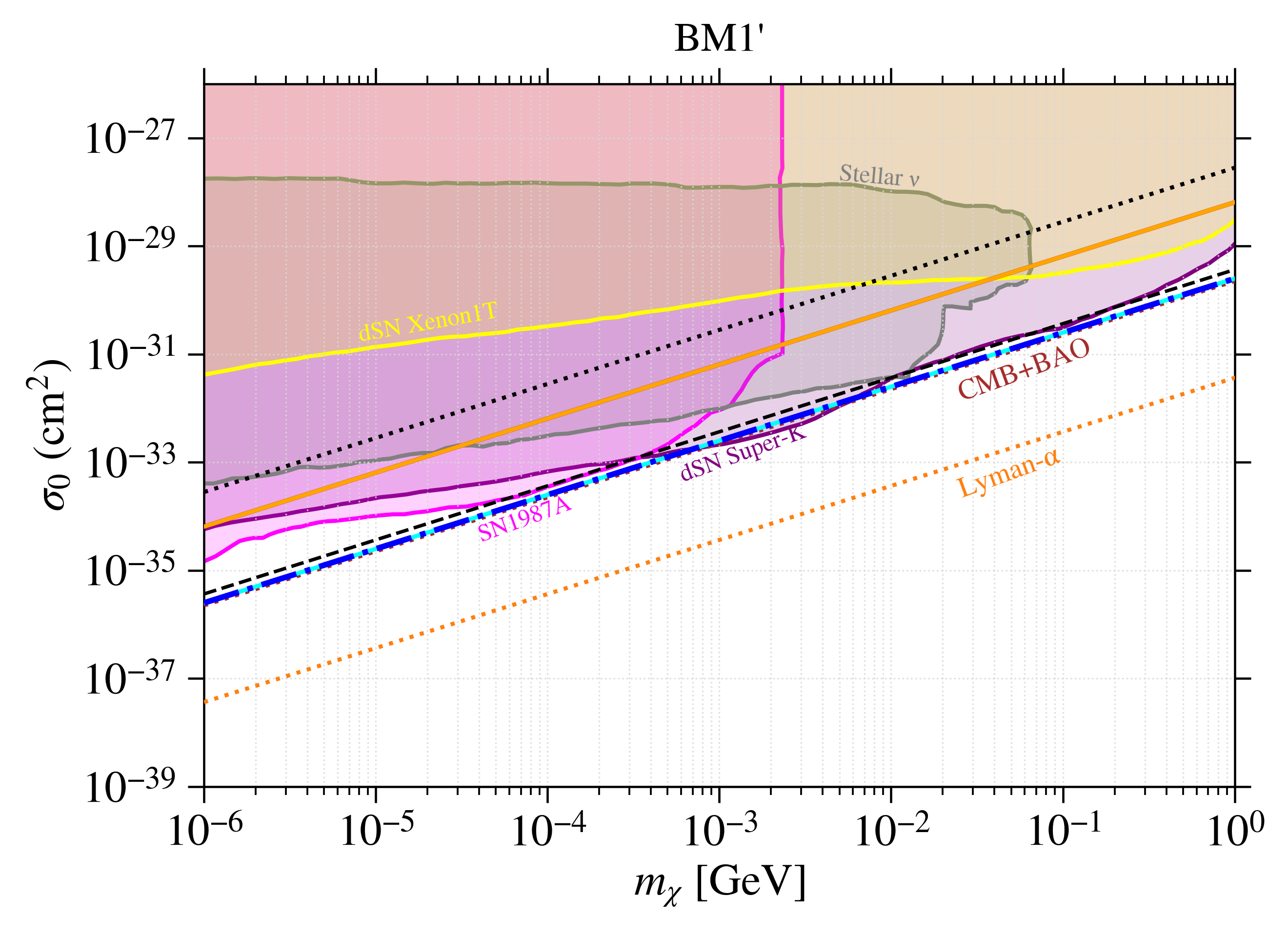}
\includegraphics[width=0.46\linewidth]{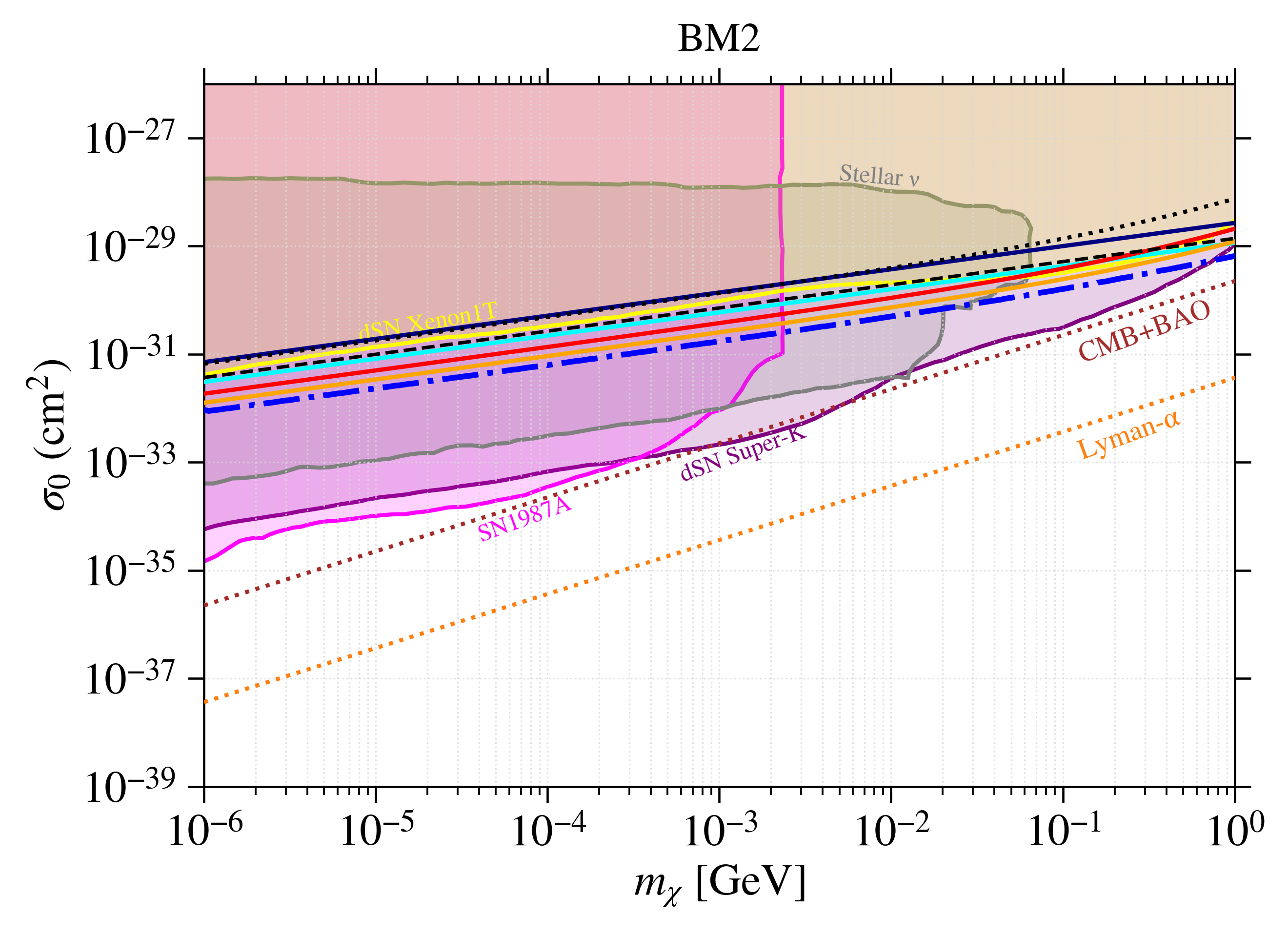}
\includegraphics[width=0.46\linewidth]{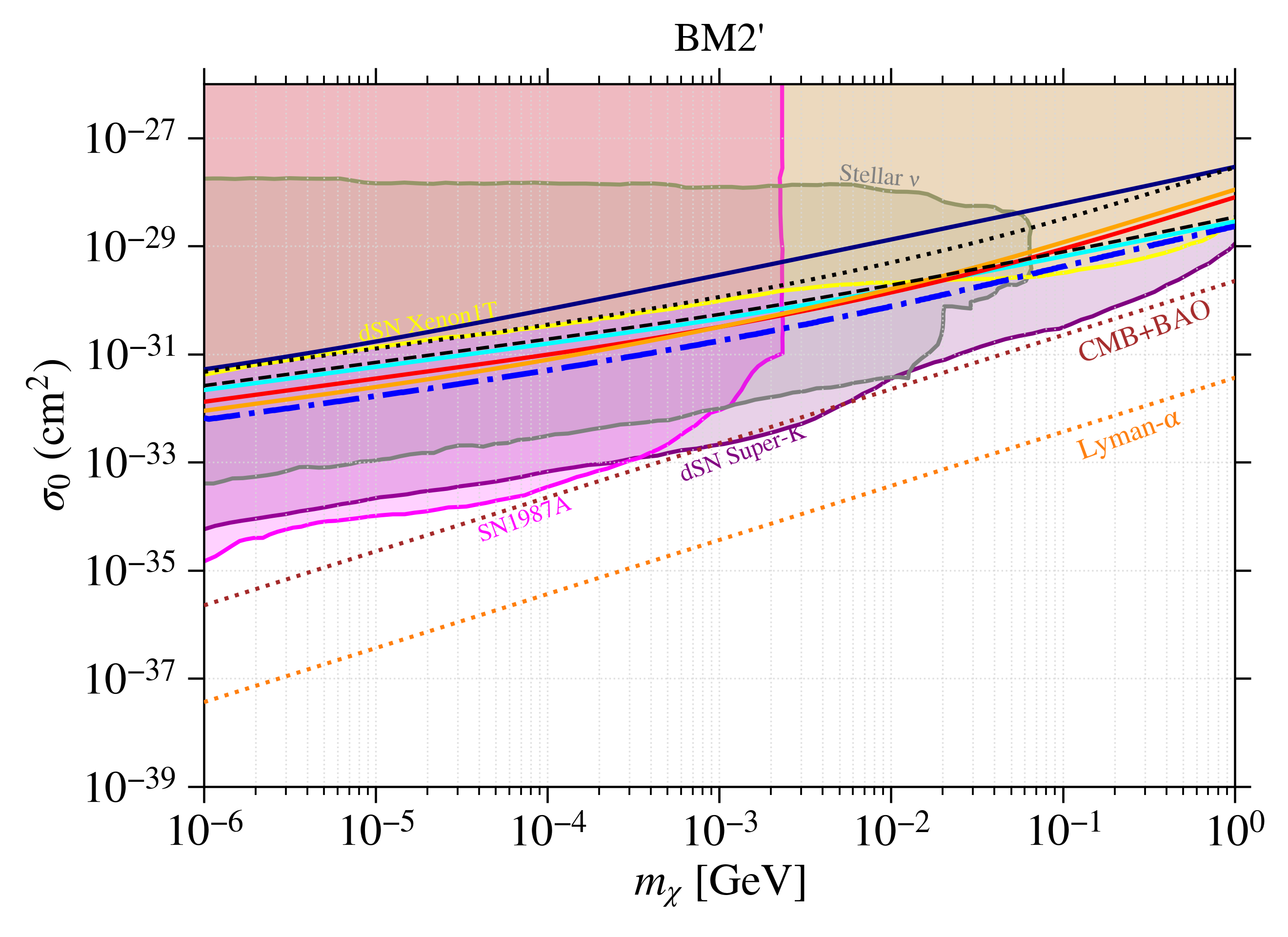}
\includegraphics[width=0.46\linewidth]{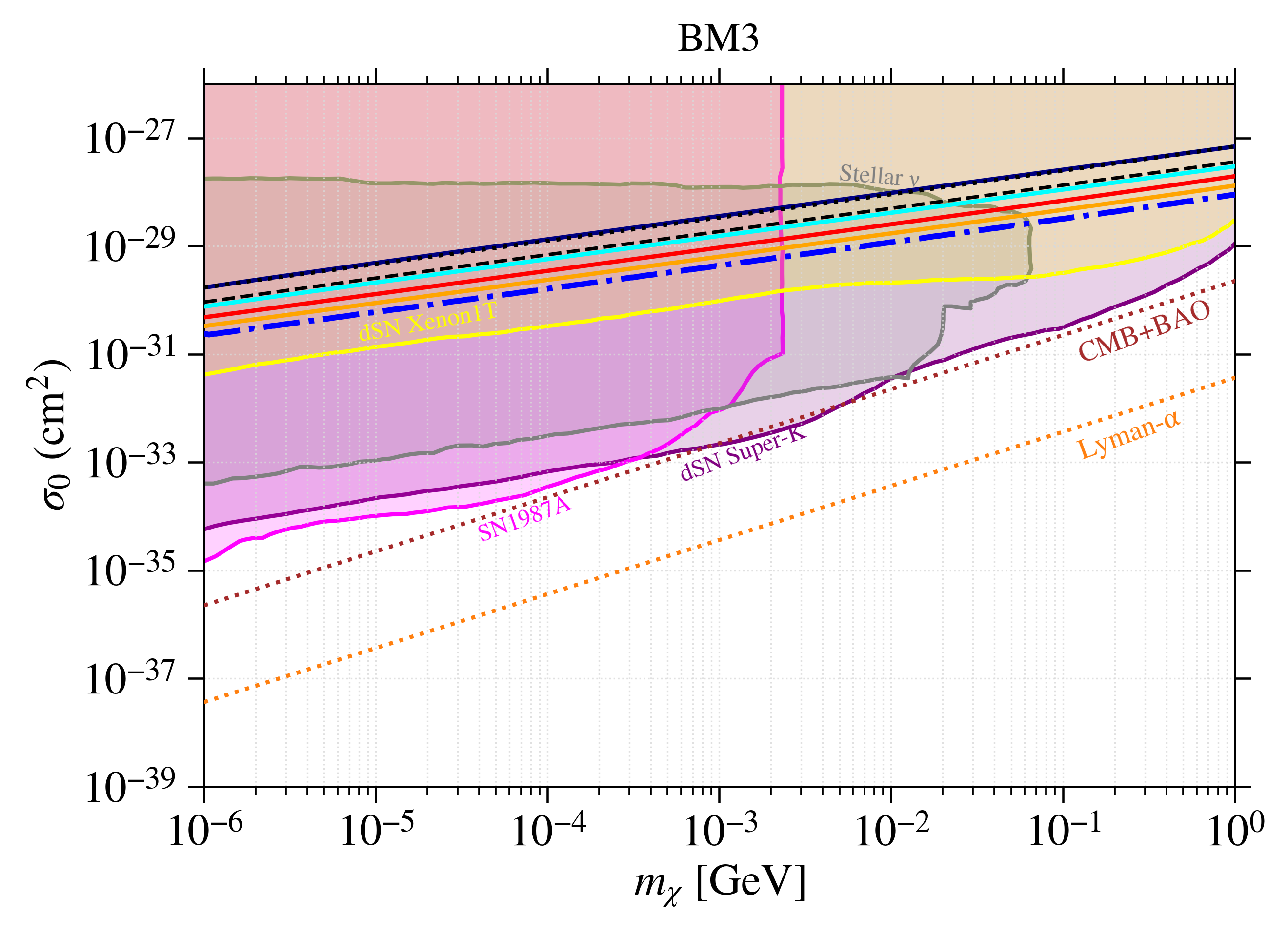}
\includegraphics[width=0.46\linewidth]{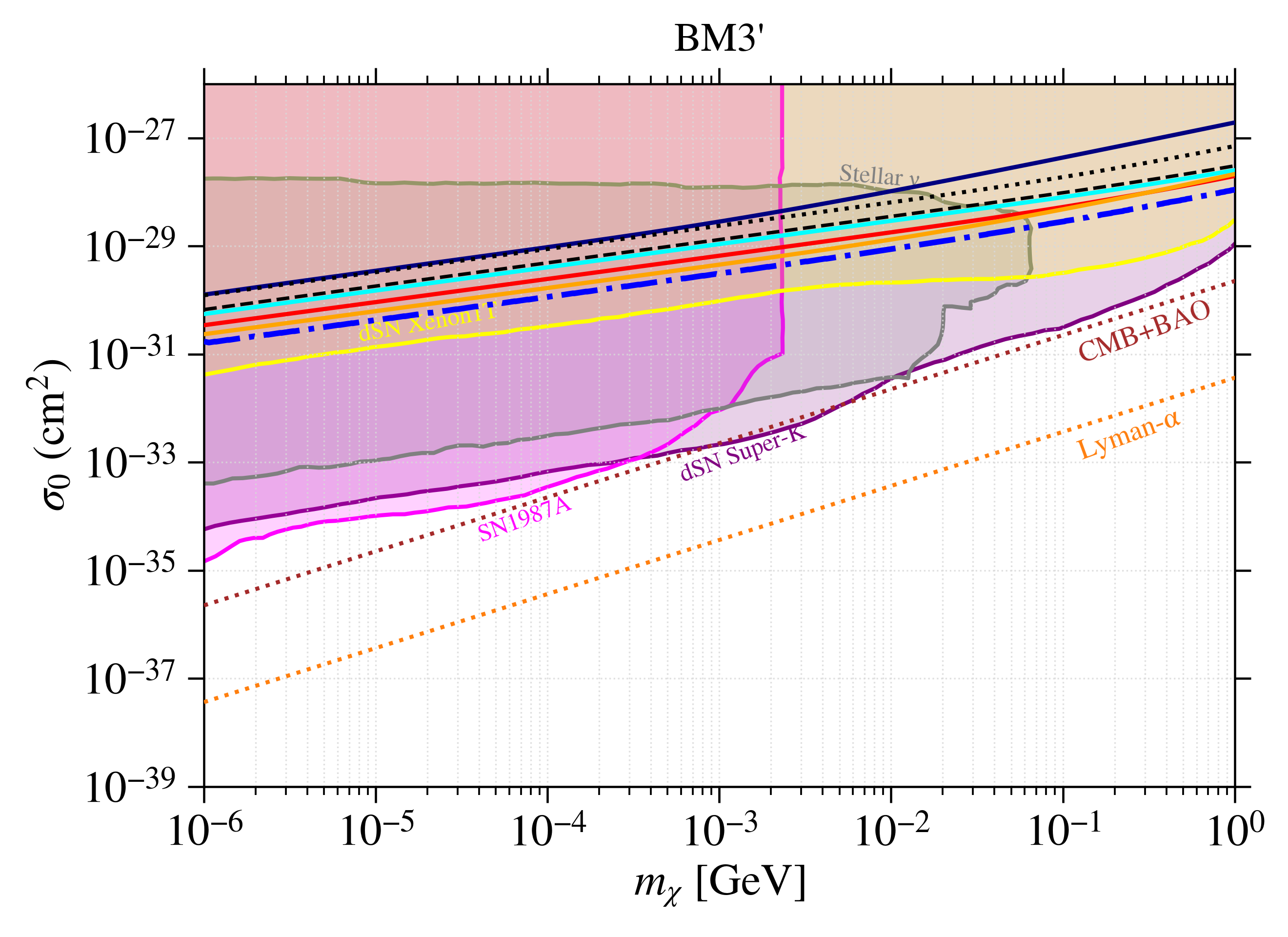}
\caption{ Constraints on $\sigma_0$ for the \textbf{constant} $\sigma_{\nu {\chi}}$ for all the benchmark models for all the sources. The blue-dotdashed line represents the bound from the stacking analysis combining the data from all four sources. The constraints obtained by Cline et al.\ for NGC 1068 \cite{Cline_2023} (black dashed) and TXS 0506+056 \cite{PhysRevLett.130.091402} (black dotted) are shown for comparisons. For context, we also overlay the model independent constraints on the DM-neutrino scattering cross-section coming from SN 1987A \cite{Lin:2022dbl} (magenta), stellar-neutrinos \cite{Jho:2021rmn} (gray solid), diffuse supernova (dSN) Xenon1T (yellow) and dSN Super-K (purple solid) \cite{Ghosh:2021vkt}, CMB+BAO \cite{Mosbech:2020ahp} (green dotted), Lyman-$\alpha$ \cite{Hooper:2021rjc} (orange dotted).}\label{fig:ConstantStackChi2}
\end{figure*}

\begin{figure*}[htbp]
\includegraphics[width=0.46\linewidth]{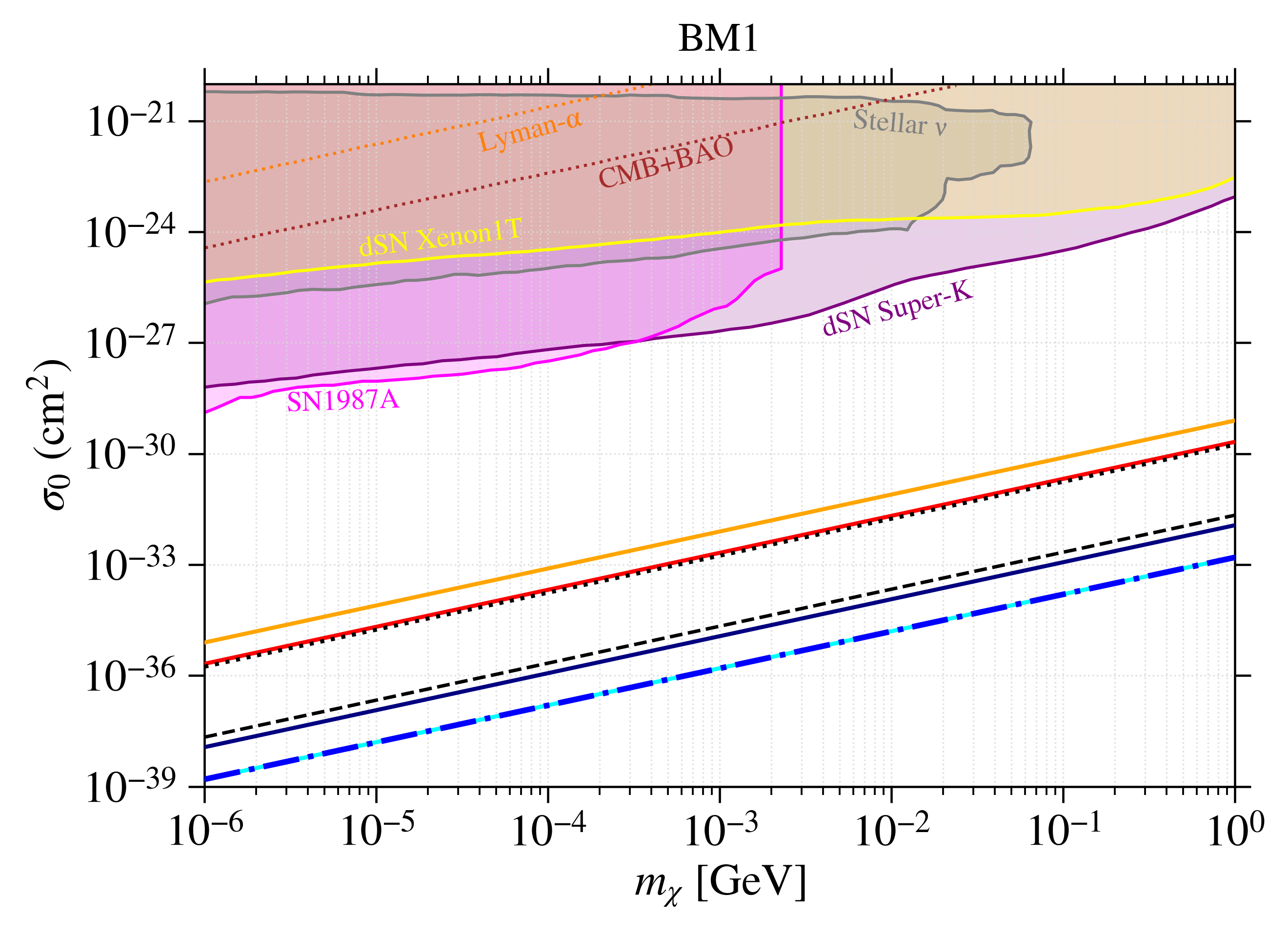}
\includegraphics[width=0.46\linewidth]{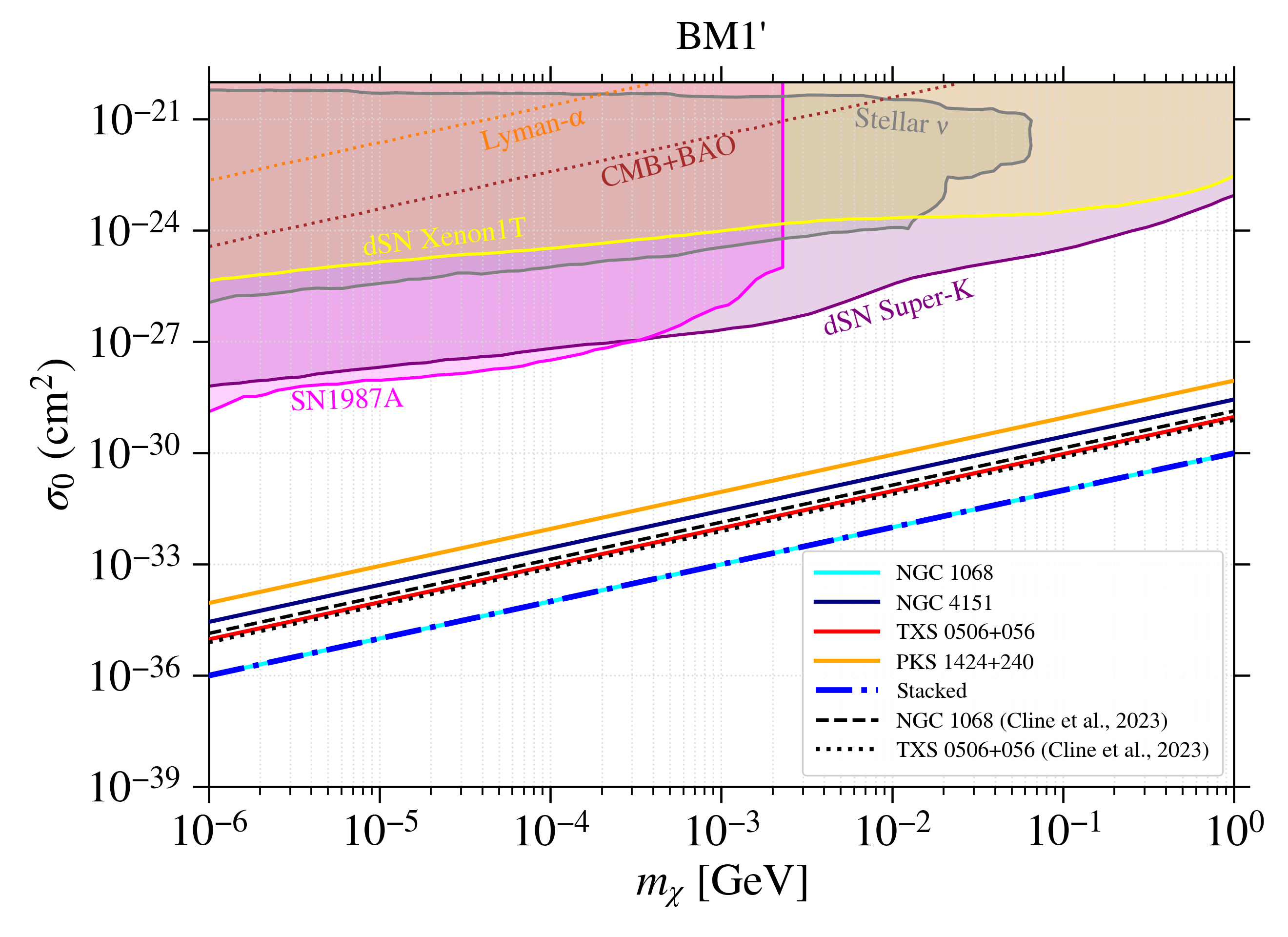}
\includegraphics[width=0.46\linewidth]{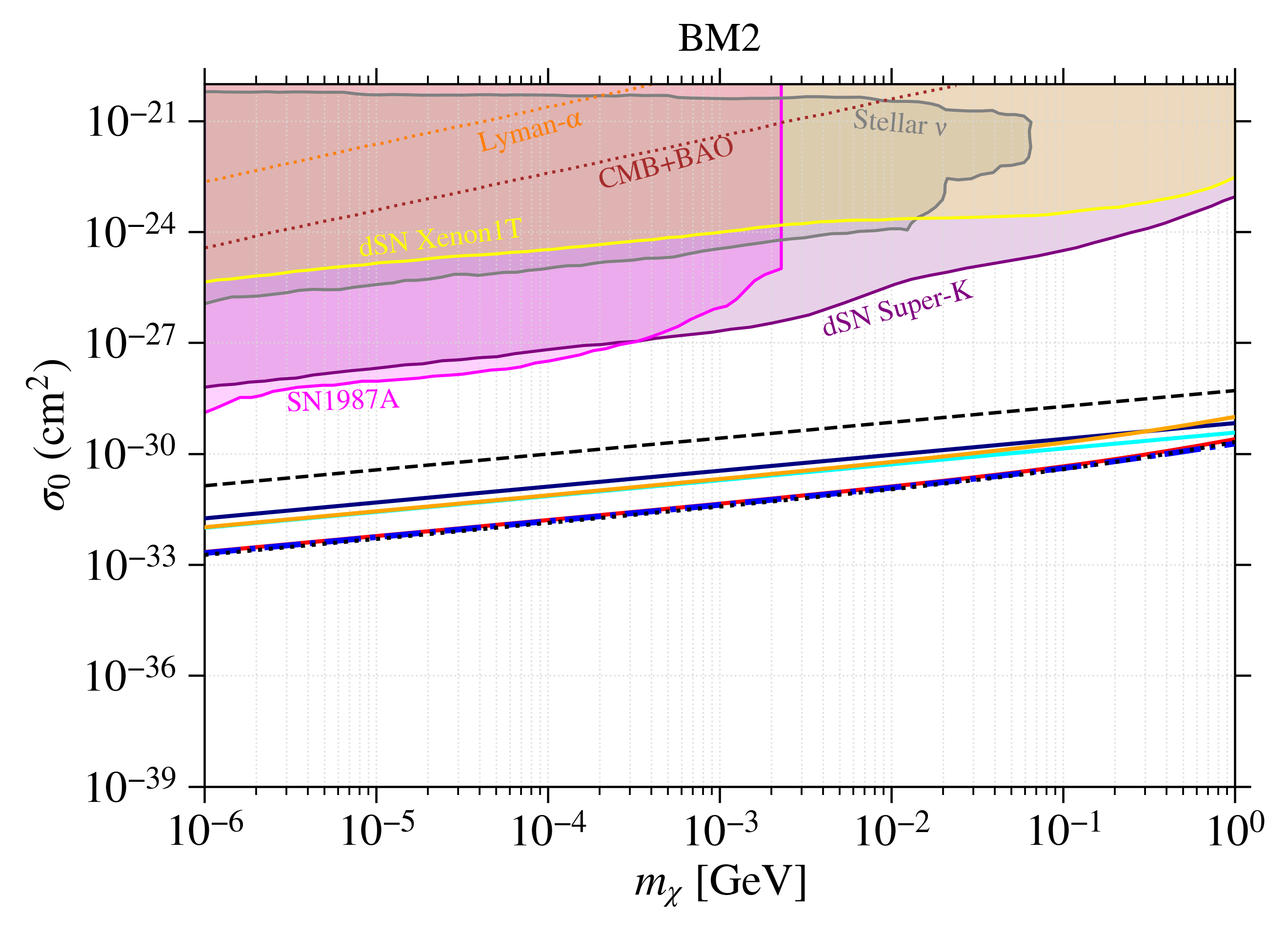}
\includegraphics[width=0.46\linewidth]{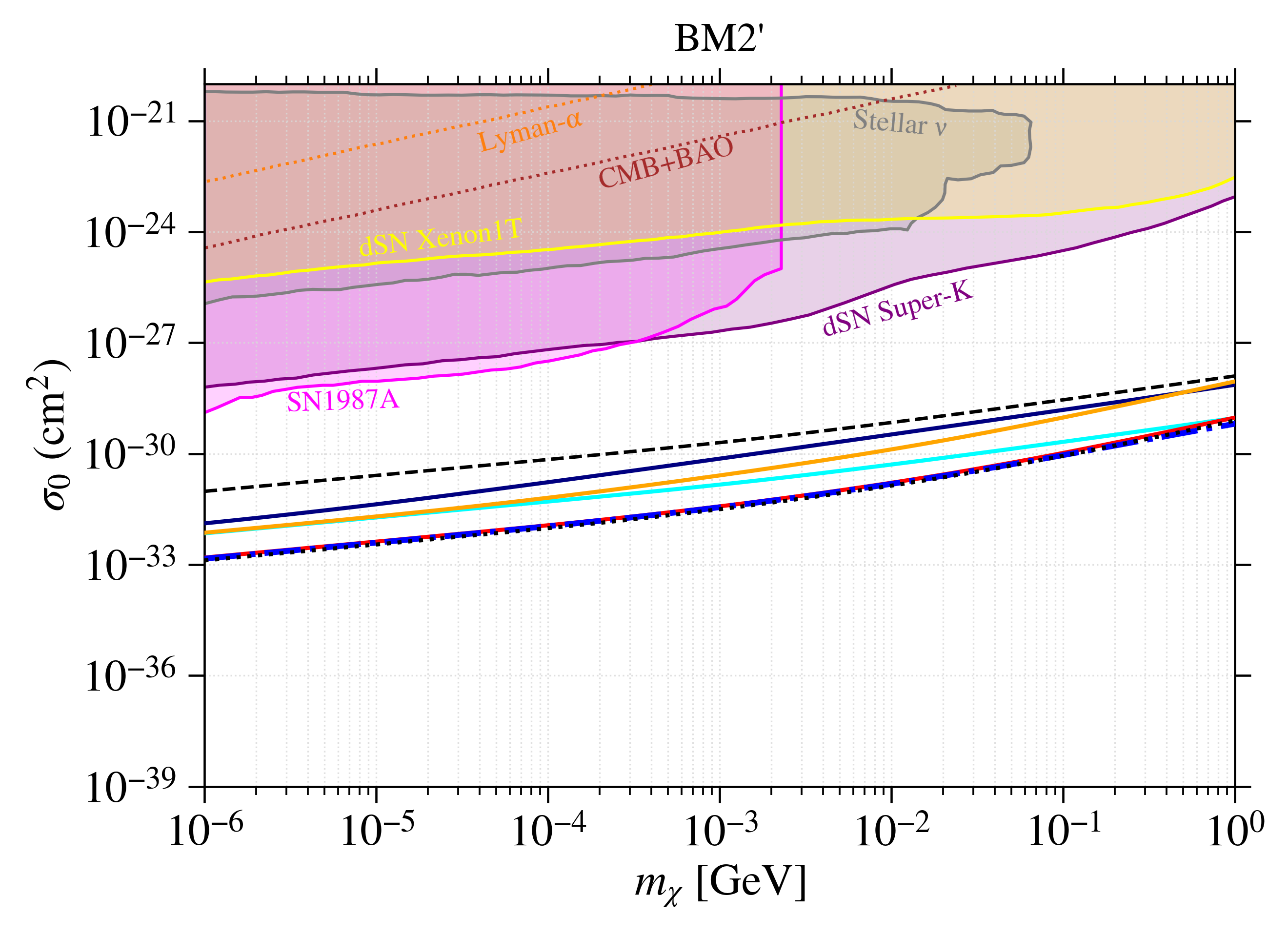}
\includegraphics[width=0.46\linewidth]{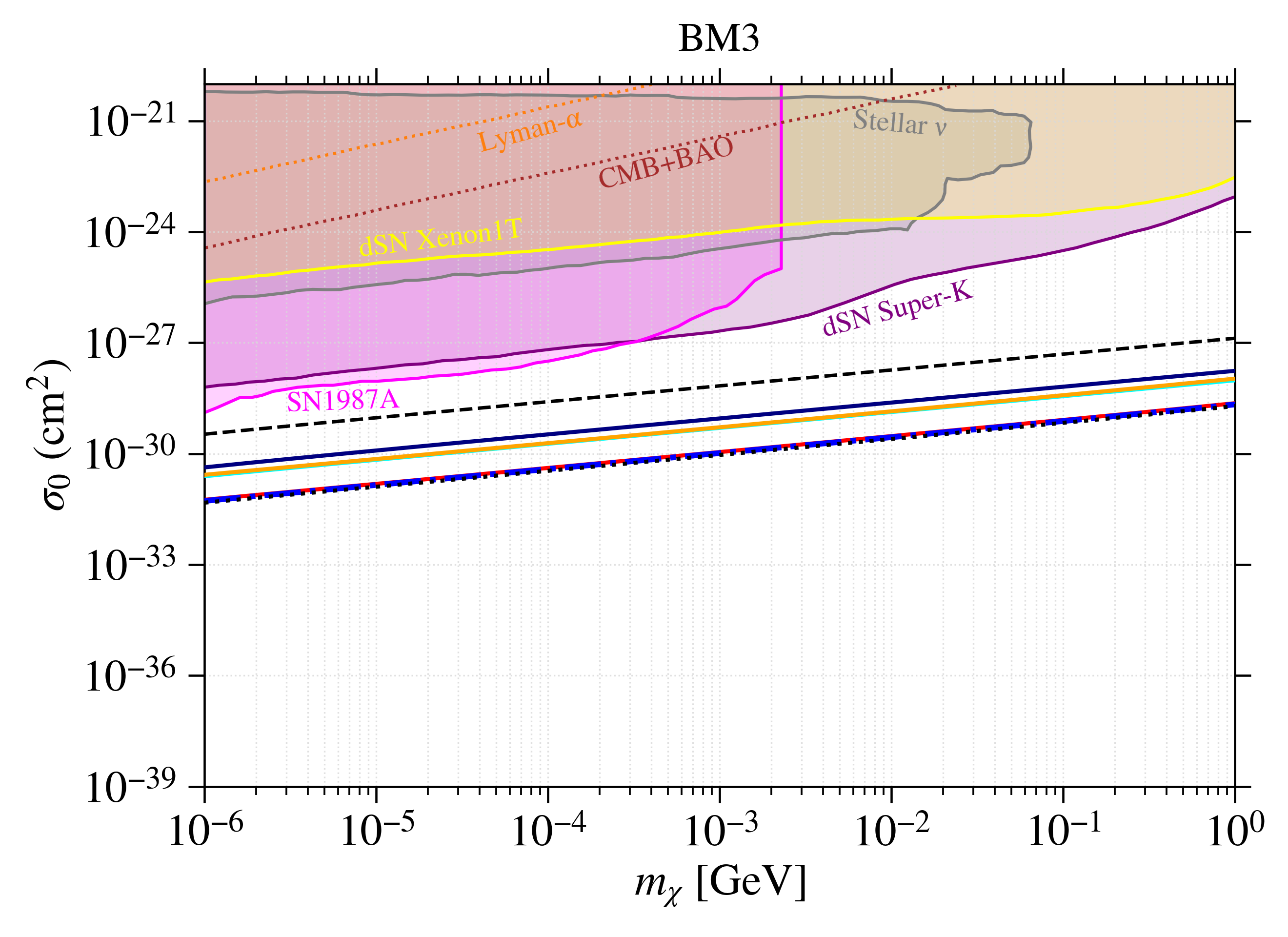}
\includegraphics[width=0.46\linewidth]{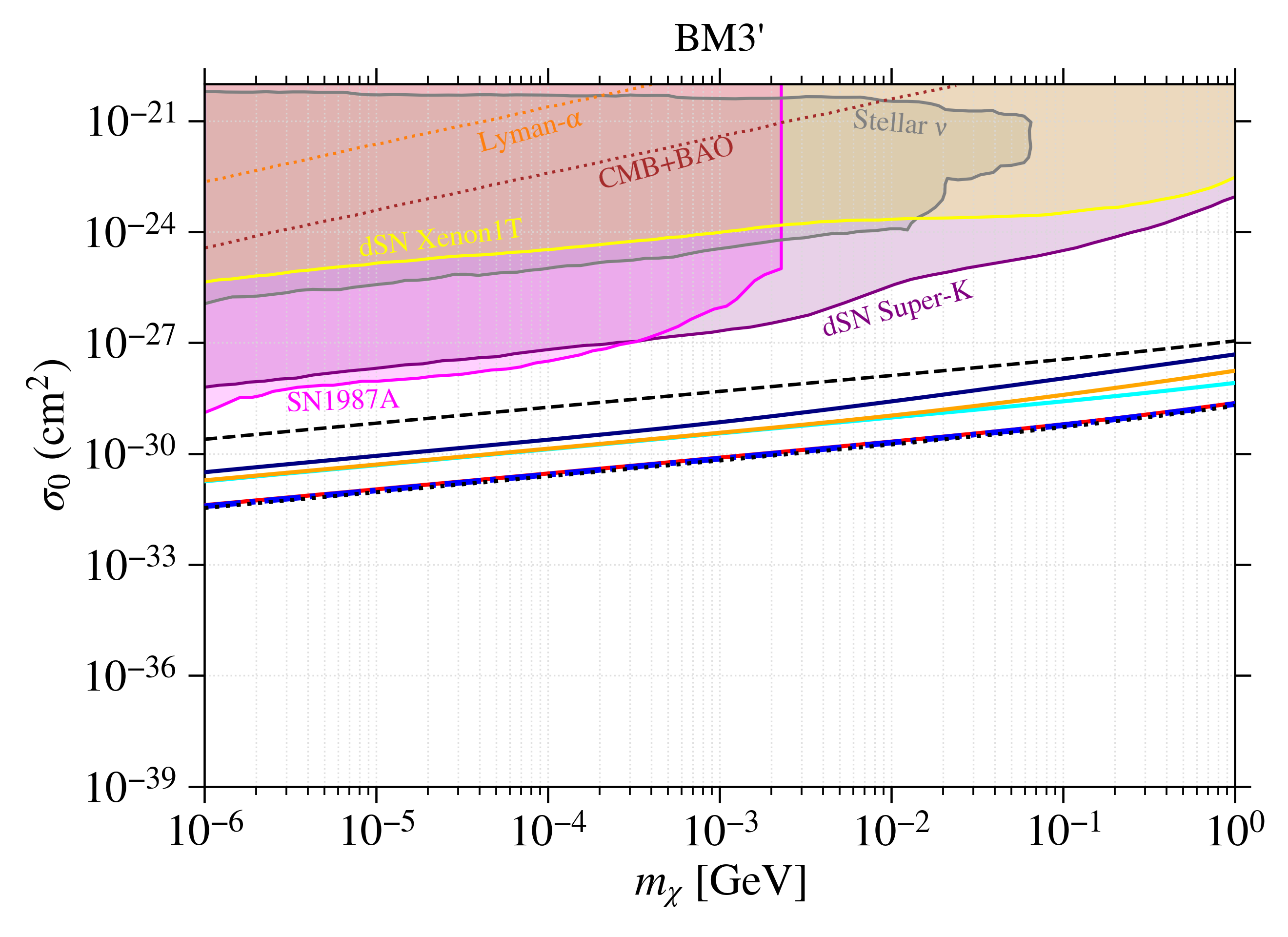}
\caption{Constraints on $\sigma_0$ for the \textbf{energy-dependent} $\sigma_{\nu \chi}$ for all the benchmark models for all the sources. The solid-blue line represents the bound from the stacking analysis combining the data from all four sources. The limits obtained previously \cite{Cline_2023,PhysRevLett.130.091402} have been rescaled to $E_0=10$ TeV for comparisons.
}\label{fig:EnergyDepStackChi2}
\end{figure*}

In this work, we consider two scenarios: (i) constant $\sigma_{\nu \chi} = \sigma_0$, and (ii) linearly energy dependent $\sigma_{\nu \chi} = \sigma_0 \frac{E_\nu}{E_0}$. In the case of constant cross-section, the second term is zero, 
hence, the attenuated flux is $\Phi = \Phi_{\nu+\bar{\nu}}~e^{-\sigma_0 \Sigma_\chi/m_\chi}$. 
However, in case of energy-dependent $\sigma_{\nu \chi}$, the energy re-distribution term cannot be neglected, and an analytic solution to the cascade equation above is highly non-trivial. Following \cite{PhysRevLett.130.091402}, we use a discretized form of this equation for each energy bin $i$ given by
\begin{equation}
\frac{d\Phi_i}{dy}
= A\!\left[
 -\,\hat{E}_i\,\Phi_i
 + \Delta x\,\ln 10
 \sum_{j=i}^{N} \hat{E}_j\,\Phi_j
\right],
\label{eq:discetizedcascade}
\end{equation}
where, $y$ is a unitless quantity that varies between $0$ and $1$, $\hat{E}_{i} = 10^{x_{i}}$, $x=\log_{10} (E_\nu/TeV)$ and
\begin{equation}
A = \frac{\Sigma_{\chi}}{m_{\chi}}\,
    \frac{\sigma_{0}}{\hat{E}_{0}}.
\label{eq:6a}
\end{equation}
For each source, we adopt a reference scale energy $E_0 = 10$~TeV, which provides a consistent basis for comparing all sources and enables a stacking analysis that combines data across the full sample.

\section{\label{sec:DataAnalysis}Data analysis and Results}

\subsection{Likelihood-based $\chi^2$ analysis}

To place constraints on the DM-neutrino scattering cross-section, from individual sources, we define a likelihood-based $\chi^2$ using Poisson statistics as follows \cite{baker1984clarification}
\begin{equation}\label{eq:chi2}
    \chi^2 = 2 \sum_i \bigg[N_i^{th} - N_i^{ex} + N_i^{ex} \ln \bigg(\frac{N_i^{ex}}{N_i^{th}}\bigg)\bigg] + 2\sum_{j\ne i}N_{j}^{th}
\end{equation}
where $i$ corresponds to the energy bins for which the experimental event number $N_i^{ex} \ne 0$ and $j$ corresponds to the energy bins for which $N_j^{ex} = 0$. The quantities $N_i^{th}$ or $N_j^{th}$ correspond to the theoretical estimate of the number of neutrino events from the flux model, after taking into account DM-neutrino scattering and standard neutrino oscillations over astrophysical distances. 
Here $N^{ex} = n_s$ as listed in Table~\ref{tab:n_s}. We perform a stacking analysis by combining $\chi^2$ from all sources (NGC~1068, TXS~0506+056, PKS~1424+240 and NGC~4151), i.e. 
\begin{eqnarray}
    \chi^2_{\rm total} = &\chi^2_{\rm (NGC~1068)} + \chi^2_{\rm (TXS~0506+056)}\\ \nonumber
    &+ \chi^2_{\rm (PKS~1424+240)} + \chi^2_{\rm (NGC~4151)},
\end{eqnarray}
which we minimize for a common $\sigma_0$ and $m_{\chi}$.

\subsection{\label{sec:Results}Results for constant cross-section}

The results of this analysis for constant  $\sigma_{\nu\chi}$ are shown in Fig. \ref{fig:ConstantStackChi2}. 
\begin{itemize}
\item We find that for benchmark models BM1 and BM1$^\prime$, for which $\langle \sigma_a v\rangle = 0$ 
and consequently the column density $\Sigma_{\chi}(m_{\chi})$ is nearly independent of the DM mass $m_{\chi}$, the strongest constraint arises from the NGC~1068 source. To understand the reason behind this result, we evaluate the $\chi^2$ given in Eq. (\ref{eq:chi2}) for the constant cross-section as follows
\begin{equation}\label{eq:simpchi}
    \chi^2 \approx n_s\big[e^{-\sigma_0 \Sigma_{\chi}/m_{\chi}}- 1 + \sigma_0 \Sigma_{\chi}/m_{\chi}\big],
\end{equation}
where $n_s$ is the total number of neutrino events over all energy bins.
From Eq.~(\ref{eq:simpchi}), the 90\% upper bound on the DM-neutrino cross-section is obtained as 
\begin{equation*}
    \sigma_0 \lesssim  c m_{\chi},
\end{equation*}
where, $c \propto \frac{1}{\sqrt{n_s}\Sigma_{\chi}}$.
The strongest bound observed for NGC~1068 is attributed to the fact that the factor $c$ calculated from the $\chi^2$ analysis is the smallest for NGC~1068, due to the observed number of events for this source being the highest. 
We also find that for more massive SMBHs, the DM density $\rho_{\chi}$ peaks at larger radii, with lower spike height, leading to a smaller column density $\Sigma_{\chi}$ along the neutrino's propagation to the Earth. Since NGC~1068 hosts the least massive SMBH among the four sources considered, it correspondingly yields the largest $\Sigma_{\chi}$. Consequently, a combination of the largest signal $n_s$ and the largest column density produces the strongest constraints for this source. As a result, NGC~1068 largely dominates the results of the stacking analysis.

\item For the benchmark models BM2, BM3, BM2$^\prime$ and BM3$^\prime$, where DM-annihilation effects are non-zero, PKS~1424+240 provides the strongest constraint for most of the range of $m_{\chi}$, as shown in Fig. \ref{fig:ConstantStackChi2}. The reason behind this is that DM spikes with higher density get depleted faster due to DM self-annihilation resulting in large decrease in $\Sigma_\chi$ as shown for NGC~1068 and NGC~4151 in Fig. \ref{fig:SigmaVsrl} and discussed in section \ref{columnden}. In comparison, the decrease in $\Sigma_\chi$ for TXS~0506+056 and PKS~1424+240 at the emission radii of neutrinos (blue shaded region in Fig.~\ref{fig:SigmaVsrl}) is smaller for most $m_\chi$ values because of their more massive black holes. Because of a larger $\Sigma_\chi$ compared to the NGCs, PKS~1424+240 and TXS~0506+056 give stronger constraints on $\sigma_0$ in these benchmark models for most of the $m_\chi$ range. 
 
\end{itemize}

\subsection{\label{sec:Results2} Results for energy dependent cross-section}

In this section, we perform a $\chi^2$ analysis for a linearly energy-dependent scattering cross-section, of the form $\sigma_{\nu \chi} = \sigma_0 \frac{E_\nu}{E_0}$. We observe the following points:
\begin{itemize}
\item For benchmark models BM1 and BM1$^\prime$ (with $\langle \sigma_a v\rangle = 0$), the strongest constraint is again obtained from NGC~1068, yielding an upper limit of roughly $\sigma_0 \le 10^{-39}$ cm$^2$. This bound is slightly more stringent than that obtained in the energy-independent (constant cross-section) scenario. 

\item For models BM2, BM3, BM2$^\prime$ and BM3$^\prime$ (with $\langle \sigma_a v\rangle \neq 0$), the strongest constraint is obtained from TXS~0506+056, in contrast to the constant cross-section case where the strongest bound comes from PKS~1424+240. 
The difference is attributed to the energy dependence of the interaction which makes the neutrino energy event distribution particularly important. Since the event distribution for TXS~0506+056 peaks at higher energy, compared to that of PKS~1424+240, it yields the most stringent constraint in the energy-dependent scenario.
The event distributions for all sources are shown in Fig. \ref{fig:eventdistribution} in the Appendix. The event distributions are calculated by dividing the neutrino energy range from 100 GeV to 10 PeV into multiple bins. For each bin, the muon neutrino and antineutrino flux, multiplied by the detector's effective area, is integrated. This process yields the signal counts, which are then plotted as the event distribution \cite{dixit2024searching}.

\item In reference \cite{PhysRevLett.130.091402}, the results are obtained for TXS~0506+056 considering the neutrino flux predicted by the lepto-hadronic model studied in \cite{Gasparyan:2021oad}, which estimated the number of events to be approximately 2. 
The 90\% CL lower limit is set to be $\sim$ 0.1 at a fixed reference energy of $E_0=290$ TeV. 
In our analysis, we perform a full statistical treatment by assuming that the neutrino flux follows the unbroken power law given in Eq. (\ref{eq:nuflux}), to derive the lower bound on the number of events over a broad energy range from 0.1 TeV - 1 PeV for each source.
With this approach, our results for TXS~0506+056 are in very good agreement with those of reported in \cite{PhysRevLett.130.091402}.

\end{itemize}

We have also summarized our results from stacking analysis in Table \ref{tab:results} for $m_{\chi}=1$ GeV across all benchmark models, considering both constant and energy-dependent $\sigma_{\nu\chi}$.

\begin{table}[htbp]
\caption{\label{tab:results}Summary of 90\% CL stacked constraints on $\sigma_0$ for constant and energy dependent cross sections for $m_\chi = 1$ GeV and $R_{em}$ values reported in Table \ref{tab:parameters}.
}
\begin{ruledtabular}
\begin{tabular}{lcc}
Benchmark & $\sigma_{\nu\chi} = \sigma_0$ & $\sigma_{\nu\chi} \propto \sigma_0 E_\nu$ \\ 
DM profile & (cm$^2$) & (cm$^2$) \\
\hline 
 BM1 & 4.76$\times 10^{-33}$ & 1.68$\times 10^{-33}$ \\ 
 BM2 & 6.23$\times 10^{-30}$ & 1.96$\times 10^{-30}$ \\ 
 BM3 & 8.81$\times 10^{-29}$ & 2.21$\times 10^{-29}$ \\ 
 BM1$^\prime$ & 2.51$\times 10^{-30}$ & 1.05$\times 10^{-30}$ \\ 
 BM2$^\prime$ & 2.10$\times 10^{-29}$ & 6.81$\times 10^{-30}$  \\
 BM3$^\prime$ & 1.07$\times 10^{-28}$ & 2.30$\times 10^{-29}$ \\ 
 NFW & 2.63$\times 10^{-25}$ & 1.19$\times 10^{-25}$ \\ 
\end{tabular}
\end{ruledtabular}

\end{table}

\subsection{\label{sec:Results2} Impact of astrophysical model uncertainties on constraints}

The constraints obtained so far are subject to astrophysical model uncertainties, especially to the uncertainty in the neutrino emission radius $R_{em}$ and the assumption regarding the initial neutrino flux, which may significantly affect the constraints on $\sigma_0$. The effect of the uncertainty in $R_{em}$ is illustrated in Fig. \ref{fig:UncertainRem} in Appendix \ref{sec:Rem_uncertainties}. The black and red bands illustrate the uncertainty in stacked constraints on $\sigma_0$ for unprimed and primed benchmark models, respectively, within the shaded $R_{em}$ range for each source as shown in Fig.~\ref{fig:SigmaVsrl}. The dot-dashed curves represent the stacked $\sigma_0$ bounds for fixed values of $R_{em}$ associated with each source, which are reported in Table \ref{tab:parameters}. The uncertainty bands are shown for both the energy-independent cross-section, $\sigma_{\nu\chi} = \sigma_0$ (left panel), and the energy-dependent cross-section, $\sigma_{\nu\chi} = \sigma_0 (E_\nu/E_0)$ (right panel). The narrow bands associated with BM2 and BM2$^\prime$, along with the nearly invisible bands for the BM3 and BM3$^\prime$ profiles in Fig. \ref{fig:UncertainRem}, can be explained using Fig. \ref{fig:SigmaVsrl}. In this figure, we observe an order of magnitude change in $\Sigma_\chi$ for BM1, while for BM1$^\prime$, $\Sigma_\chi$ changes by a few factors within the shaded region. Also, there are negligible alterations in $\Sigma_\chi$ for BM2, BM3, BM2$^\prime$, and BM3$^\prime$. This situation is clearly visible in Fig. \ref{fig:UncertainRem}.
We can see that the constraints obtained in our analysis can vary by one order of magnitude depending on the neutrino-emission radius.

We also considered the initial neutrino flux at the source to be a broken power-law (BP) given by
\begin{equation}\label{eq:nufluxbroken}
  \Phi = \begin{cases}
\Phi_0\left(\frac{E_{\nu}}{E_b} \right)^{-\Gamma_1}, & E\leq E_b,\\
\Phi_0\left(\frac{E_{\nu}}{E_b} \right)^{-\Gamma_2}, & E > E_b.
\end{cases} ,
\end{equation}
where, we considered $\Phi_0 = \Phi_{\rm ref}$ and $\Gamma_1 \equiv \Gamma$ defined in Eq. (\ref{eq:nuflux}) and left $E_{\rm b}$ and $\Gamma_2$ as free parameters. We performed a likelihood analysis of fitting the BP model to IceCube data following our previous work \cite{dixit2024searching} where we fitted an UP to the same data. The change in test statistics, $TS=-2\log \mathcal{L}$, between the two flux models is $\Delta TS = TS_{\rm BP} - T_{\rm UP}=$ 0.49 (NGC 1068), 0.43 (TXS 0506+056), 0.06 (PKS 1424+240) and 0.2 (NGC 4151). Since the improvement in the $TS$ values for BP are not significant, the simpler UP model sufficiently describes IceCube data for each source. Nevertheless, we performed an event distribution calculation for the BP case, a similar analysis shown in Appendix \ref{sec:nu_events} for the UP case. We found that constraints on $\sigma_{\nu\chi}$ do not change in any significant manner for the energy-dependent cross-section but there is a weakening of the constraint in the energy-dependent case. These results are shown in Fig.~\ref{fig:BrokenUnbroken} of Appendix \ref{sec:UP_BP_uncertainties} for the BM1 profile, which provides the most stringent constraints on $\sigma_0$. The weakened constraint for the BP spectrum is, however, within the uncertainty band due to emission radius uncertainty shown in Fig.~\ref{fig:UncertainRem}.

\section{Model Interpretation
\label{sec:model}}
In this section, we will interpret our results in the parameter space of a well-motivated model of DM with significant coupling to neutrinos.
We consider a scenario in which the SM is extended to a gauged and anomaly-free $U(1)_{L_{\mu}-L_{\tau}}$ group \cite{He:1990pn, He:1991qd}.
Spontaneous symmetry breaking leads to a massive gauge boson $Z^{\prime}$ that couples primarily to $2^{\rm nd}$ and $3^{\rm rd}$ generation SM leptons ($\mu$, $\tau$, $\nu_{\mu}$, $\nu_{\tau}$) and their anti-particles.

The Lagrangian of interactions is given by
\begin{align}
\mathcal{L} \supset \frac{m_{Z^{\prime}}^{2}}{2} Z_{\mu}^{\prime} Z^{\prime \mu} + Z_{\mu}^{\prime} (g_{f} \mathcal{J}_{f}^{\mu} + \varepsilon e \mathcal{J}_{EM}^{\mu} + g_{\chi} \mathcal{J}_{\chi}^{\mu}) , 
\end{align}
where $m_{Z^{\prime}}$ denotes the gauge boson mass, and $g_{f} \equiv Q_{\mu-\tau} g_{\mu-\tau}$ is the gauge coupling to the SM $2^{\rm nd}$ and $3^{\rm rd}$ generation leptons. Throughout this work, we assume a unit charge, $Q_{\mu-\tau} = 1$.
The $L_{\mu}-L_{\tau}$ current is given by 
\begin{align}
\mathcal{J}_{f}^{\mu} &= \bar{\mu} \gamma^{\mu} \mu + \bar{\nu}_{\mu} \gamma^{\mu} P_{L}\nu_{\mu} - \bar{\tau} \gamma^{\mu} \tau - \bar{\nu}_{\tau} \gamma^{\mu} P_{L}\nu_{\tau},  
\end{align}
where $P_{L} = \frac{1}{2}(1 - \gamma^{5})$ is the left handed chirality operator.
The kinetic mixing $\varepsilon$, between $Z^{\prime}$ and SM gauge bosons is generated at the 1-loop level and is given by \cite{Araki:2017wyg, Bernal:2025szh}
\begin{align}
\varepsilon &= -\frac{e g_{\mu-\tau}}{12 \pi^{2}} {\rm log} \left(\frac{m_{\tau}^{2}}{m_{\mu}^{2}}\right) \nonumber \\
&\simeq -\frac{g_{\mu-\tau}}{70}, 
\end{align}
with the electromagnetic current for all SM fermions represented by $\mathcal{J}_{EM}^{\mu} = \bar{f}\gamma^{\mu} f$. 
Although kinetic mixing induces a coupling of all other SM fermions to the $Z^{\prime}$, these interactions are subdominant.
We consider a DM $\chi$ charged under this symmetry, with $g_{\chi} \equiv Q_{\mu-\tau}^{\chi} g_{\mu-\tau}$ is the coupling of $Z^{\prime}$ to DM.
We assume that DM is vector-like under this gauge symmetry such that the charge ($Q_{\mu-\tau}^{\chi}$) can differ from unity \cite{Kahn:2018cqs, Buckley:2022btu}.
For illustration, we focus on two benchmark DM scenarios (pseudo-Dirac fermion and complex scalar), in which the corresponding current is given by
\begin{equation}
\mathcal{J}_{\chi}^{\mu} = \left\{\begin{array}{lc} \bar{\chi}_{1} \gamma^{\mu} \chi_{2} & {\rm Pseudo-Dirac} \\
i \chi^{*} \partial^{\mu} \chi + c.c &  {\rm Complex ~Scalar}.
\end{array} \right.
\label{eq:dm_models}
\end{equation}

\begin{figure*}[htbp]
\includegraphics[width=0.49\linewidth]{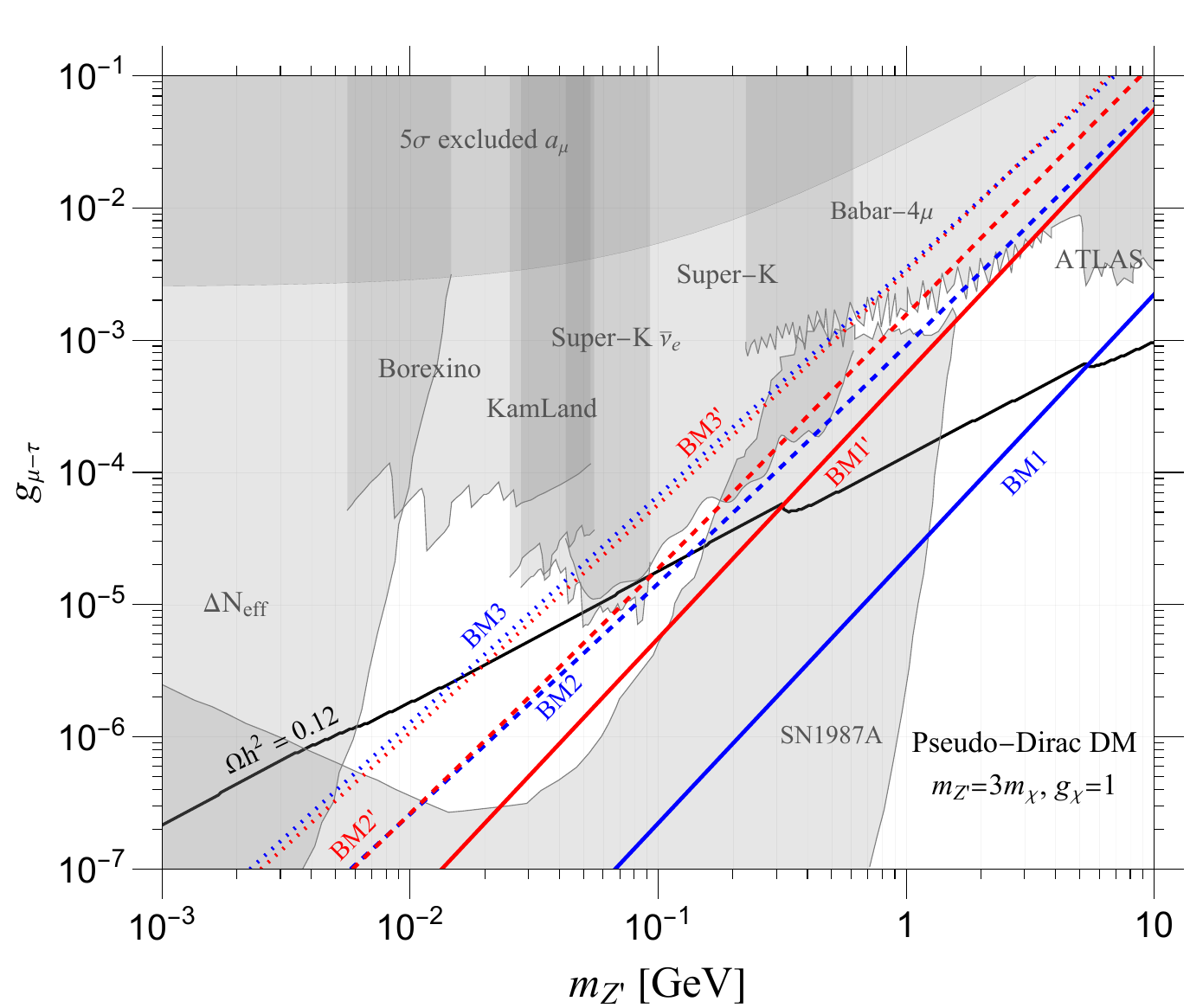}
\includegraphics[width=0.49\linewidth]{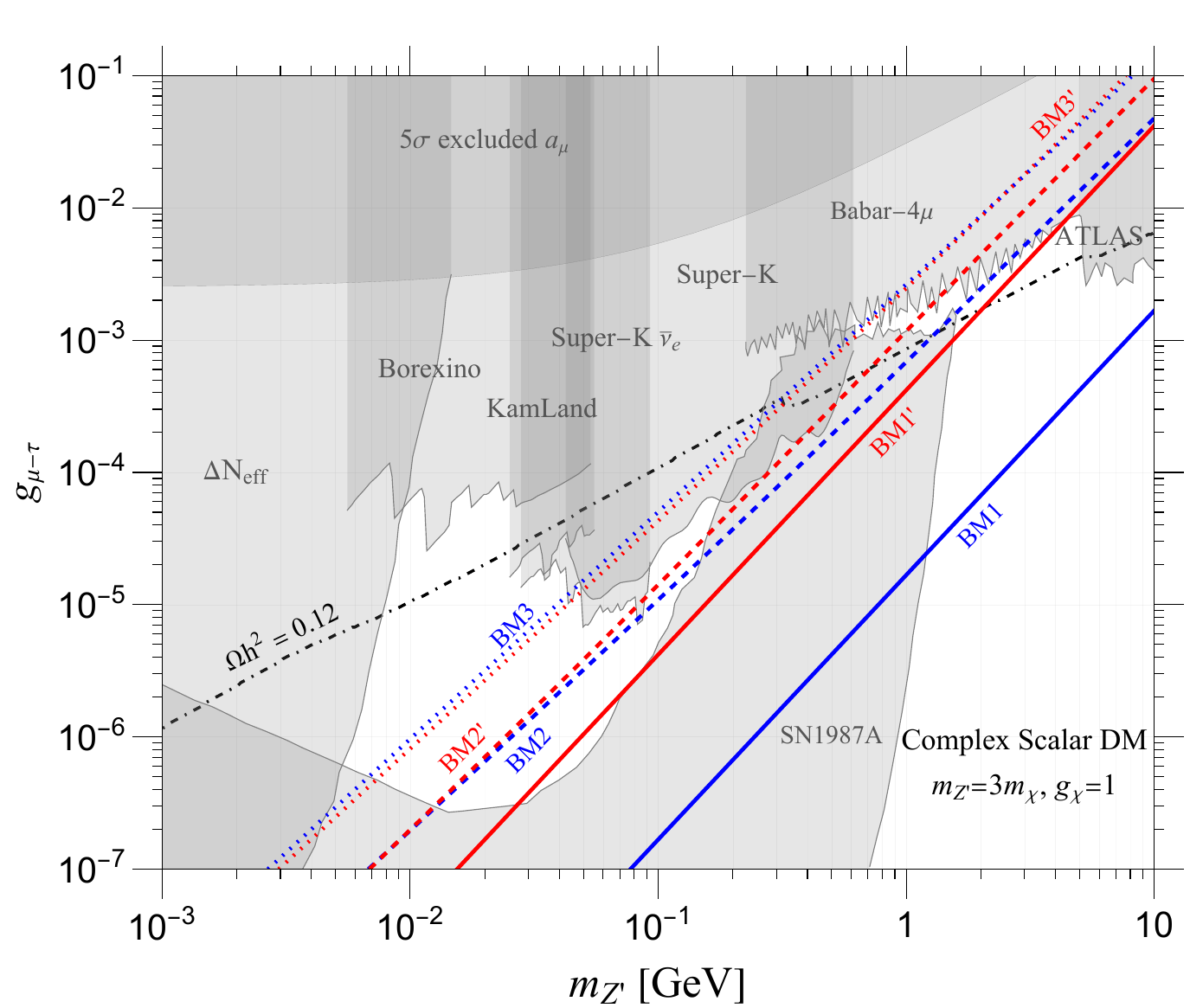}
\caption{Limits on the model parameters for energy-dependent scattering cross sections are shown for pseudo-Dirac dark matter (left panel) and complex scalar dark matter (right panel). In each panel, the black solid and dot-dashed curves indicate parameter values that reproduce the observed dark matter relic abundance with $\Omega h^{2} = 0.12$ \cite{Planck:2018vyg}. The colored curves show the constraints obtained from our stacking analysis for the different SMBH spike density benchmark models considered. The gray shaded regions denote existing bounds on the parameter space from complementary probes, including constraints from dark matter annihilation into neutrinos \cite{Buckley:2022btu}, as well as limits from early-Universe cosmology, astrophysical observations, and terrestrial accelerator searches for new gauge bosons in this class of models \cite{Bernal:2025szh}.
} 
\label{fig:gmunuVsmzl}
\end{figure*}

\begin{figure*}[htbp]
\includegraphics[width=0.55\linewidth]{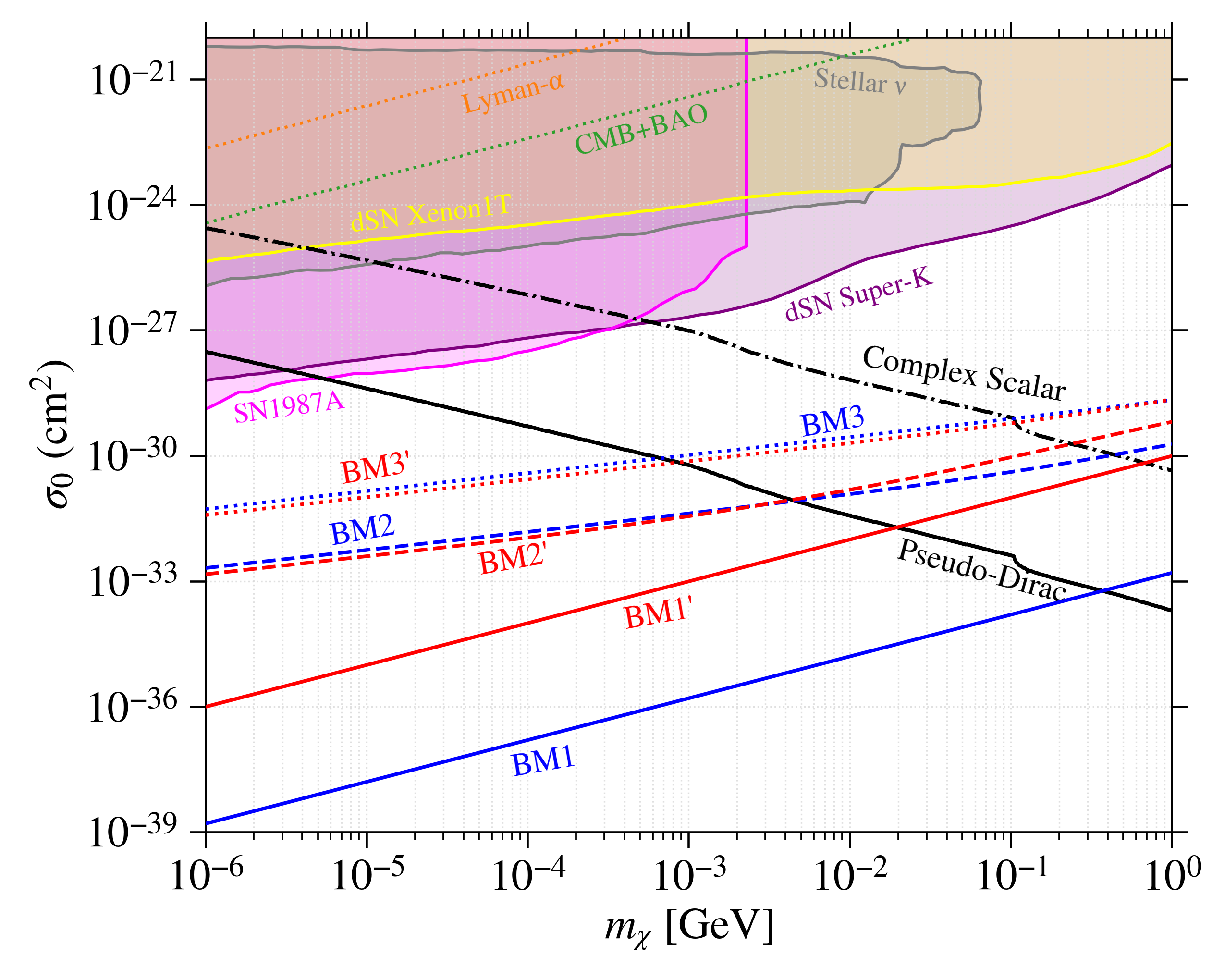}
\caption{Limits on the dark matter-neutrino scattering cross-section vs dark matter mass. We include the freeze-out relic density in our two dark matter scenarios of interest. We overlay our stacking analysis results for each benchmark model spike density model.
} 
\label{fig:ModelCroSecVsmchi}
\end{figure*}

\subsection{Dark Matter Relic Density}
We consider thermal relic DM whose present-day abundance is set by the freeze-out mechanism.
We focus on the regime $m_{Z^{\prime}} > m_{\chi}$, with $m_{\chi}$ the DM mass, such that the dominant annihilation channel is $\bar{\chi} \chi \rightarrow \bar{f} f$.
In both model scenarios considered above, the corresponding annihilation cross-section (in terms of Mandelstam s), for $\chi$ into $f= \mu,\tau$, is given by
\begin{align}
\sigma_{ann}(s) = \frac{g_{\chi}^{2}g_{\mu-\tau}^{2}}{12 \pi s} \frac{\sqrt{s - 4 m_{f}^{2}}(s + 2 m_{f}^{2})(s + 2 m_{\chi}^{2})}{(s - 4m_{\chi}^{2})[(s - m_{Z^{\prime}}^{2})^{2} + m_{Z^{\prime}}^{2} \Gamma_{Z^{\prime}}^{2}]}, 
\end{align}
where $m_{f}$ is the SM fermion mass, and the total $Z^{\prime}$ decay width to $\mu$ and $\tau$ is given by
\begin{align}
\Gamma_{Z^{\prime}} = \frac{g_{\mu-\tau}^{2} m_{Z^{\prime}}}{12 \pi} \left( 1 + \frac{2 m_{f}^{2}}{m_{Z^{\prime}}^{2}}\right)\sqrt{1 - \frac{4 m_{f}^{2}}{m_{Z^{\prime}}^{2}}}.
\end{align}
The decay width into neutrinos is
$\Gamma_{Z^{\prime}} \simeq \frac{g_{\mu-\tau}^{2} m_{Z^{\prime}}}{24 \pi}$.
In the non-relativistic limit, the thermally averaged annihilation cross-section can be parametrized as  $\langle \sigma_{a}v \rangle \equiv \sigma_{a} \, x^{-n}$\cite{Wells:1994qy, Kahn:2018cqs, Buckley:2022btu},
where $x\equiv \frac{m_{\chi}}{T}$ is a parameter representing the temperature. The parameter n characterizes the dominant partial wave of the annihilation process: $n=0$ for s-wave annihilation (as in the pseudo-Dirac scenario) and $n=1$ for p-wave annihilation (as in the complex scalar case).
Hence, the thermally averaged annihilation cross-section for both scenarios is generalized as
\begin{align}
\sigma_{a}^{f} = \frac{g_{\mu-\tau}^{2} g_{\chi}^{2}}{k \pi m_{\chi}^{2}}\frac{(2 + m_{f}^{2}/m_{\chi}^{2})\sqrt{1 - m_{f}^{2}/m_{\chi}^{2}}}{[(4 - m_{Z^{\prime}}^{2}/m_{\chi}^{2})^{2} + m_{Z^{\prime}}^{2} \Gamma_{Z^{\prime}}^{2}/m_{\chi}^{4}]}, 
\end{align}
and into neutrinos;
\begin{align}
\sigma_{a}^{\nu} = \frac{g_{\mu-\tau}^{2} g_{\chi}^{2}}{k \pi m_{\chi}^{2}}\frac{1}{[(4 - m_{Z^{\prime}}^{2}/m_{\chi}^{2})^{2} + m_{Z^{\prime}}^{2} \Gamma_{Z^{\prime}}^{2}/m_{\chi}^{4}]}, 
\end{align}
with $\sigma_{a} = \sigma_{a}^{f} + \sigma_{a}^{\nu}$.
We follow the freeze-out formalism in \cite{ Buckley:2022btu} and obtain the analytic approximation of the relic abundance of $\chi$ as
\begin{align}
\Omega_{\chi} h^{2} = 8.77\times 10^{-11} \frac{l(n+1)x_{f}^{n+1}{\rm GeV^{-2}}}{(g_{*,s}/\sqrt{g_{*}})\sigma_{a}},  
\end{align}
where $l=2$ if DM has an antiparticle and $l=1$ if it is its own antiparticle.
$g_{*,s}$ and $g_{*}$ are the entropic degrees of freedom and the relativistic degrees of freedom, respectively, with $x_{f}$ the freeze-out temperature.
For the models described above, the factors $(n,k,l) = (0,2,2)$ and $(1,12,2)$ for the pseudo-dirac and complex scalar models, respectively \footnote{Adopting the precise treatment of \cite{Steigman:2012nb}, for our benchmark parameters, has a negligible impact on our results}.

\subsection{Dark matter - Neutrino Scattering}
Neutrinos with energy $E_{\nu}$ can scatter off ambient DM, which we take to be at rest in the Galactic frame. These interactions can modify the propagation of high-energy neutrinos leading to potentially observable effects in neutrino telescopes.
Within the DM scenarios introduced above, the DM-neutrino scattering proceeds via exchange of the new gauge boson $Z^{\prime}$. At leading order, the elastic scattering cross-section for pseudo-Dirac DM is given by 
\begin{widetext}
\begin{equation}
\begin{aligned}
\sigma_{\chi \nu} = \frac{g_{\mu-\tau}^{2} g_{\chi}^{2}}{32\pi E_{\nu}^{2} m_{\chi}^{2}} \Bigg[ (m_{Z^{\prime}}^{2} + m_{\chi}^{2} + 2 E_{\nu} m_{\chi})\,{\rm log} \left(\frac{m_{Z^{\prime}}^{2}(2 E_{\nu} + m_{\chi})}{m_{\chi}(4 E_{\nu}^{2} + m_{Z^{\prime}}^{2}) + 2E_{\nu}m_{Z^{\prime}}^{2}}\right)\\ + 4 E_{\nu}^{2}\left(1 + \frac{m_{\chi}^{2}}{m_{Z^{\prime}}^{2}}-\frac{2 E_{\nu}(4 E_{\nu}^{2}m_{\chi} + E_{\nu}(m_{\chi}^{2} + 2 m_{Z^{\prime}}^{2}) + m_{\chi} m_{Z^{\prime}}^{2})}{(2 E_{\nu} + m_{\chi})(m_{\chi}(4 E_{\nu}^{2} + m_{Z^{\prime}}^{2}) + 2 E_{\nu} m_{Z^{\prime}}^{2})} \right) \Bigg].
\end{aligned}
\end{equation}
\end{widetext}

For complex scalar DM, the cross-section is given by
\begin{widetext}
\begin{equation}
\begin{aligned}
\sigma_{\chi \nu} = \frac{g_{\mu-\tau}^{2} g_{\chi}^{2}}{16\pi E_{\nu}^{2} m_{\chi}} \left[ \frac{4 E_{\nu}^{2} m_{\chi}}{m_{Z^{\prime}}^{2}} + (m_{\chi} + 2 E_{\nu})\,
{\rm log} \left( \frac{m_{Z^{\prime}}^{2}(2 E_{\nu} + m_{\chi})}{m_{\chi}(4 E_{\nu}^{2} + m_{Z^{\prime}}^{2})+ 2E_{\nu}m_{Z^{\prime}}^{2}}\right) \right].
\end{aligned}
\end{equation}
\end{widetext}
Both cross-sections were obtained following the description in \cite{Arguelles:2017atb} and capture the full energy dependence of high energy neutrinos.
However, they exhibit limiting behaviors, which are expressed as 
\begin{align}
\sigma_{\chi \nu} = \frac{g_{\mu-\tau}^{2} g_{\chi}^{2}}{\kappa \pi \, m_{Z^{\prime}}^{2}} \left\{\begin{array}{lc} 1  & E_{\nu} \gg m_{Z^{\prime}}/m_{\chi} \\
m_{\chi} E_{\nu}/m_{Z^{\prime}}^{2} &  E_{\nu} \ll m_{Z^{\prime}}/m_{\chi},
\end{array} \right.
\label{eq:sigma_lim}
\end{align}
where $\kappa = 8$ for the pseudo-dirac DM and $\kappa = 4$ for the complex scalar DM \footnote{We have verified that our calculations match the results of \cite{Arguelles:2017atb} up to a prefactor of $\frac{1}{2}$, which comes from the $P_{L}$ operator in the $Z^{\prime}$-$\nu$ coupling.}.
Equation (\ref{eq:sigma_lim}) justifies our choices of $\sigma_{\chi 
\nu} = \sigma_{0}$ or $\sigma_{\chi 
\nu} \sim \sigma_{0} \, E_{\nu}$, as assumed throughout the study and illustrated in Figs.~\ref{fig:ConstantStackChi2} and \ref{fig:EnergyDepStackChi2}.\\

In Fig.~\ref{fig:gmunuVsmzl}, we show the results of our stacking analysis overlaid on the $U(1)_{L_{\mu}-L_{\tau}}$ model parameter space, plotting the gauge coupling $g_{\mu -\tau}$ vs gauge boson mass $m_{Z^{\prime}}$.
On the left and right panels we showcase the results of the pseudo-Dirac fermion and complex scalar DM, respectively, assuming $m_{Z^{\prime}} = 3 m_{\chi}$ and $g_{\chi} = 1$. The color lines represent our constraints incorporating the different spike density profiles after stacking the results from TXS 0506+056, PKS 1424+240, NGC 4151 and NGC 1068 (the blue dot-dashed line in each panel of Figs.~\ref{fig:ConstantStackChi2} and \ref{fig:EnergyDepStackChi2}). 
The solid blue and red lines represent the BM1 and $\rm BM1^{\prime}$ models, respectively. The dashed lines correspond to the BM2 and $\rm BM2^{\prime}$ models, while the dotted lines represent the BM3 and $\rm BM3^{\prime}$ spike density models, respectively.

The gray shaded regions indicate constraints on the model arising from various experimental and observational searches. The regions associated with the neutrino experiments (i.e. Borexino, KamLand, Super-K) correspond to parameter space in which galactic center DM annihilates into neutrinos; see \cite{Buckley:2022btu} for further details.
Constraints from early-universe cosmology, astrophysical observations and terrestrial probes, including the latest results from the muon g-2 measurements, are discussed in  \cite{Bernal:2025szh}.\\

In this model, DM can also interact with electrons or nucleons, via kinetic mixing. However, constraints from direct detection experiments on this space are relatively weak (due to the suppressed kinetic mixing) and are therefore not shown. For brevity, we also omit other existing model constraints that are subdominant compared to those presented here.

Finally, in the left panel, the black solid line denotes the DM relic density predicted by the pseudo-Dirac fermion model. In the right panel, the dot-dashed line shows the dark matter relic abundance in the complex scalar model.
Within the parameter space shown, the IceCube stacking analysis assuming the BM1 spike density profile yields the most stringent constraints on the gauge coupling. In particular, it constrains a substantial portion of the relic density in the pseudo-Dirac fermion scenario and effectively the entirety of the relic density-favored region in the complex scalar case.

Fig.~\ref{fig:ModelCroSecVsmchi} illustrates the impact of our stacking analysis on the DM parameter space, compared to other model-independent bounds by recasting the thermal relic density in the $\sigma_{0}$ vs $m_{\chi}$ plane. The black solid and dot-dashed curves correspond to the pseudo-Dirac and complex scalar models respectively, with $m_{Z^{\prime}} = 3 m_{\chi}$ and $g_{\chi} = 1$. For illustration, we present the results in the limit $E_{\nu} \ll m_{Z^{\prime}}/m_{\chi}$ corresponding to the energy-dependent scattering cross-section scenario shown in Fig.~\ref{fig:EnergyDepStackChi2}.
The color shaded regions indicate the model-independent bounds as discussed above. Overall, we find that our particle physics model-independent stacking analysis provides stringent constraints on the DM parameter space for the chosen benchmark parameters.

\section{\label{sec:Conclusions}Discussion \& Conclusions}
We analyzed high energy astrophysical neutrino data from AGN and blazars, as observed by IceCube, to place constraints on the DM-neutrino scattering cross-section. In contrast to the current literature, where individual sources were analyzed and constraints derived from these, we perform a statistically robust stacking analysis in which we combine the results from 4 individual sources of extragalactic high energy neutrinos, namely TXS 0506+056, PKS 1424+240, NGC 4151 and NGC 1068.

Consistent with the literature, we perform our analysis for both a constant cross-section and a cross-section linearly dependent on the neutrino energy. The strongest constraints observed in our analysis are $ \sigma_{0} \lesssim 8 \times 10^{-39}$ cm$^2$ and $ \sigma_{0} \lesssim 10^{-39}$ cm$^2$ in case of constant and energy-dependent cross sections, respectively, at 90\% CL for 1 keV DM mass, which is derived from the stacking analysis for BM1 spike density profile under the assumption over the initial neutrino flux following an unbroken power-law in Eq. (\ref{eq:nuflux}). However, BM1-type steep spikes are likely to occur less frequently in many AGNs, especially in scenarios involving merger activity. Additionally, the derived constraints roughly scale with the assumed column density $\Sigma_\chi$ as given in Table \ref{tab:parameters}. Thus, it is imperative to interpret the strongest bounds as fundamentally reliant on spike survival. While a broken power-law model of the initial neutrino flux is not statistically favored, adopting such a model somewhat weakens our constraint for the energy-dependent cross-section. Furthermore, our constraints can be weakened by up to an order of magnitude if the uncertainty on the neutrino emission radius is considered.

For context, it is useful to compare our results with existing bounds in the literature. Previous cosmological analyses based on Milky Way satellite galaxy counts have placed upper limits of $\sigma_{\rm DM-\nu} \lesssim 4 \times 10^{-34}\,{\rm cm}^2 (m_{\rm DM}/{\rm GeV})$ at 95\% CL for a constant DM–neutrino scattering cross section, corresponding to relic neutrinos with characteristic energies of order $10^{-4}\,\mathrm{eV}$ \cite{Akita:2023yga}. These constraints arise from the impact of DM–neutrino interactions on small-scale structure formation in the early Universe. In contrast, our analysis probes the same interaction using high-energy astrophysical neutrinos detected by IceCube, extending up to TeV energies, and provides the most stringent limits to date in this high-energy regime. For completeness, we also note that a comparatively weaker bound, of order $10^{-22}\,{\rm cm}^2\,{\rm GeV}^{-1}$, has been reported using the highest-energy event observed by KM3NeT, where only propagation through the Milky Way dark matter halo was considered \cite{Bertolez-Martinez:2025trs}. Related constraints using the same event, including extragalactic and possible host-halo dark matter contributions under a blazar-origin hypothesis, have also been investigated \cite{Mondol:2025uuw}.

Our results from the stacking analysis, in case of constant cross-section, are dominated by NGC~1068 due to the largest observed number of events and the expected large column density for this source. The benchmark model BM1, however, does not take into account effects such as gravitational scattering of DM with surrounding stars or self-annihilation of DM. These effects have been incorporated in other benchmark models BM1$^\prime$, BM2/3 and so on, and we found that our constraints on the DM-neutrino scattering cross-section are weaker for those models. 
Additionally, for those benchmark models, we found that the most stringent constraints on the DM-neutrino scattering cross-section come from sources with heavier SMBHs, making TXS~0506+056 and PKS~1424+240 particularly significant in this context. 

In the energy-dependent cross-section scenario, the most stringent constraint arises for spike profile BM1 from NGC~1068,
which is stronger than that derived in the energy-independent (constant cross-section) case. In contrast, for DM density profiles surrounding SMBHs with a nonzero $\langle \sigma_a v \rangle$ ($i.e.$, BM2, $\rm BM2^{\prime}$, BM3, $\rm BM3^{\prime}$), the strongest constraint instead comes from TXS~0506+056. This can be attributed to the detection of the highest-energy neutrinos from this source (with energies up to $\approx$290 TeV), which enhances experimental sensitivity in the energy-dependent regime.

In our analysis of model interpretations, we discussed the anomaly-free $U(1)_{L_\mu-L_\tau}$ model and presented our results from a stacking analysis conducted in the model's parameter space. This model considers the scenario of pseudo-Dirac and complex scalar DM, where we established the most stringent constraints on the coupling and mediator mass parameters.
A recent analysis of DM–neutrino interactions has examined scalar and fermionic DM coupled through various mediators, considering laboratory, cosmological, and astrophysical constraints, where they excluded many previously accepted scenarios \cite{Dev:2025tdv}. On the other hand, our study focuses on a well-motivated anomaly-free model, deriving high-energy constraints in its parameter space from IceCube observations at TeV energies.

In conclusion, the constraints obtained in this study on DM-neutrino scattering cross-sections from individual sources are significantly enhanced when we conduct a stacking analysis for each assumed DM density profile around SMBHs. Collectively, these findings provide the most stringent constraints on the DM-neutrino scattering cross-section in this context to date.

\begin{acknowledgments}
We thank Aaron Vincent, Gabrijela Zaharijas and Matteo Puel for helpful discussions. K.D. and S.R. were partially supported by a National Research Foundation (NRF) of South Africa grant facilitated through the National Institute of Theoretical and Computational Sciences (NITheCS). GM is supported by the Natural Sciences and Engineering Research Council of Canada (NSERC). During the completion of this work, GM was also supported by the University of California Irvine, School of Physical Sciences visiting fellowship as well as the Aspen Center for Physics, which is supported by the National Science Foundation grant PHY-2210452.  
\end{acknowledgments}

\onecolumngrid

\appendix

\setcounter{figure}{0}
\renewcommand{\thefigure}{\thesection\arabic{figure}}

\section{DM column density}
\label{sec:column}

\begin{figure*}[htbp]
\includegraphics[width=0.4\linewidth]{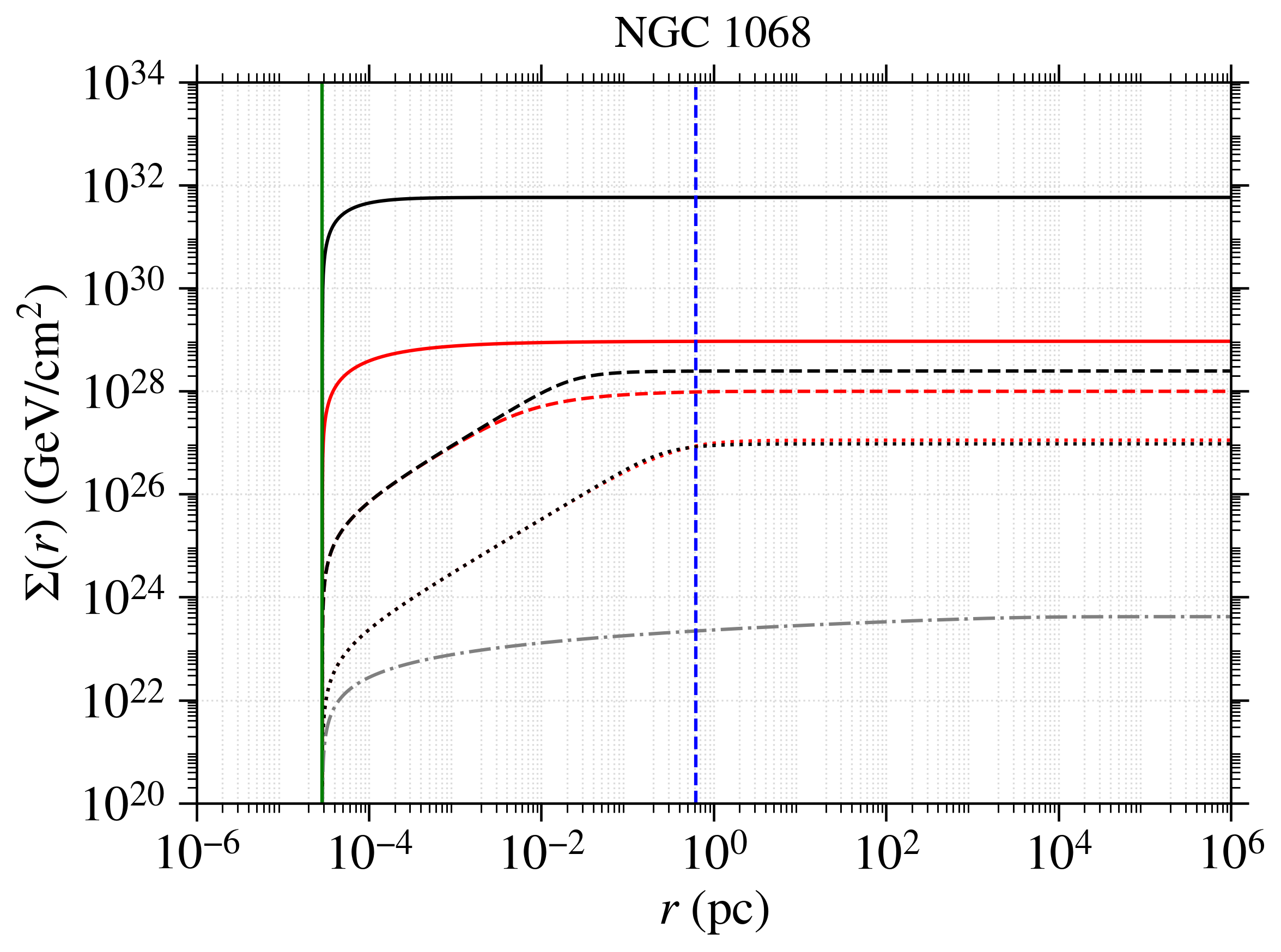}
\includegraphics[width=0.4\linewidth]{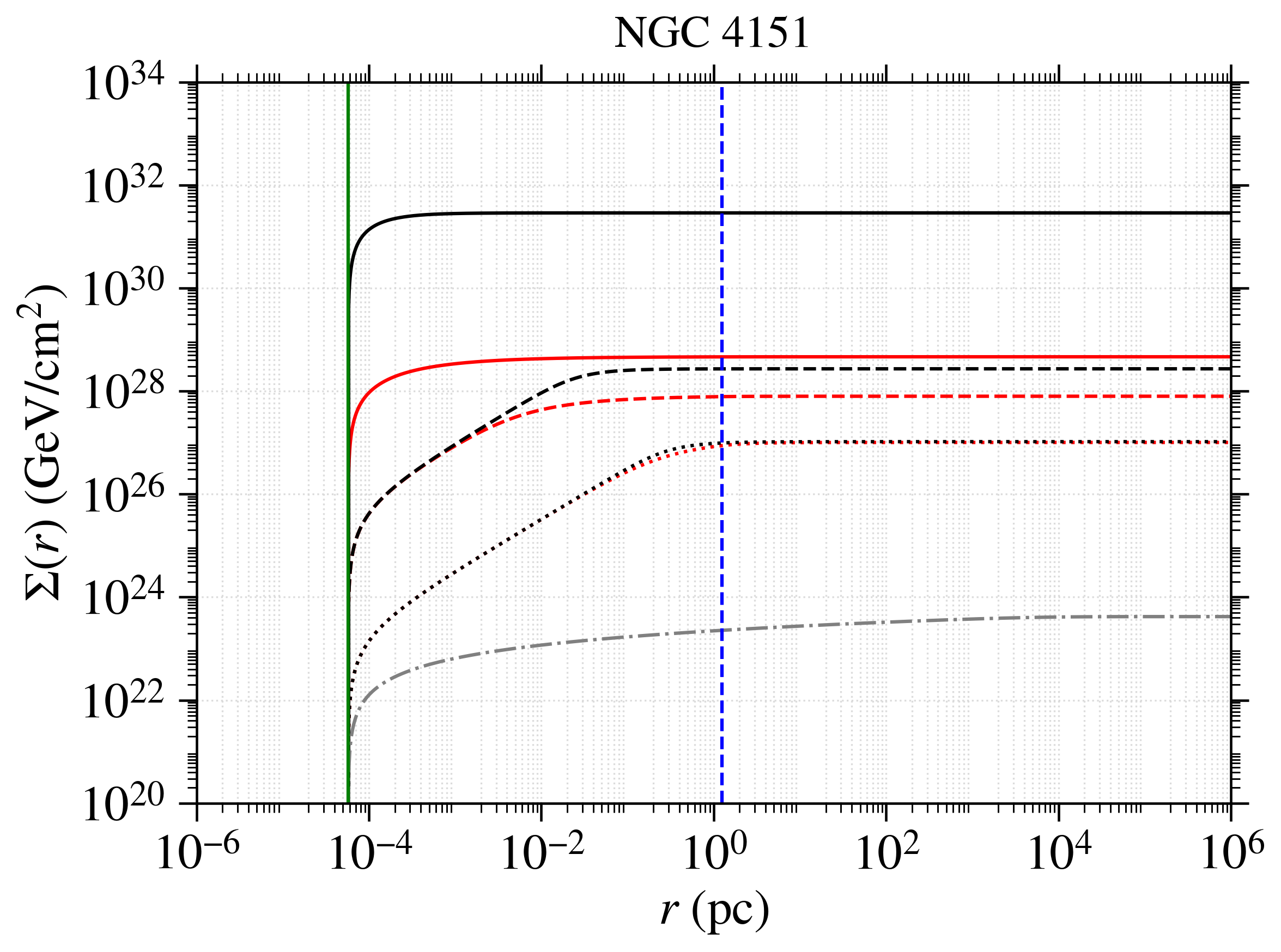}
\includegraphics[width=0.4\linewidth]{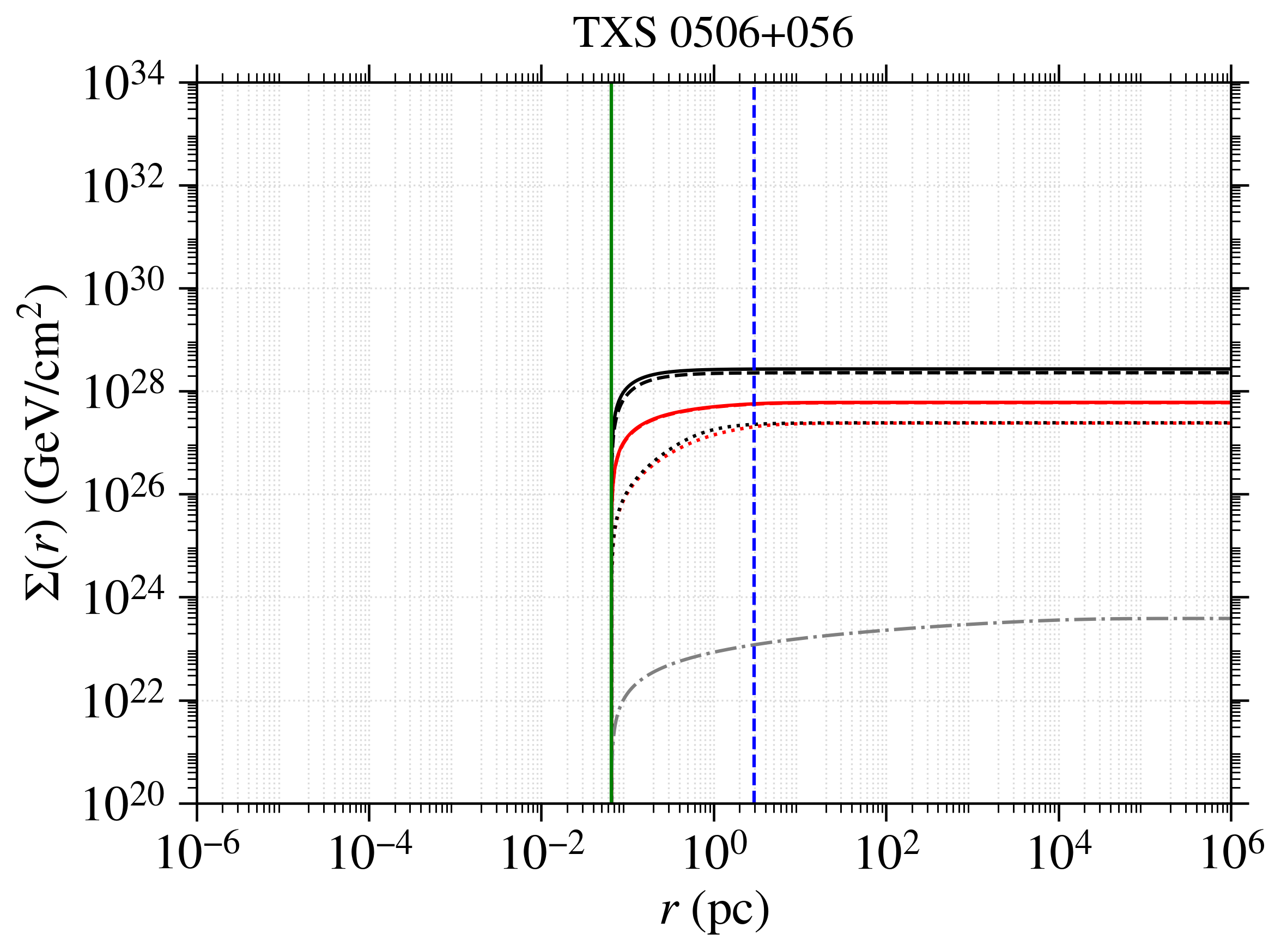}
\includegraphics[width=0.4\linewidth]{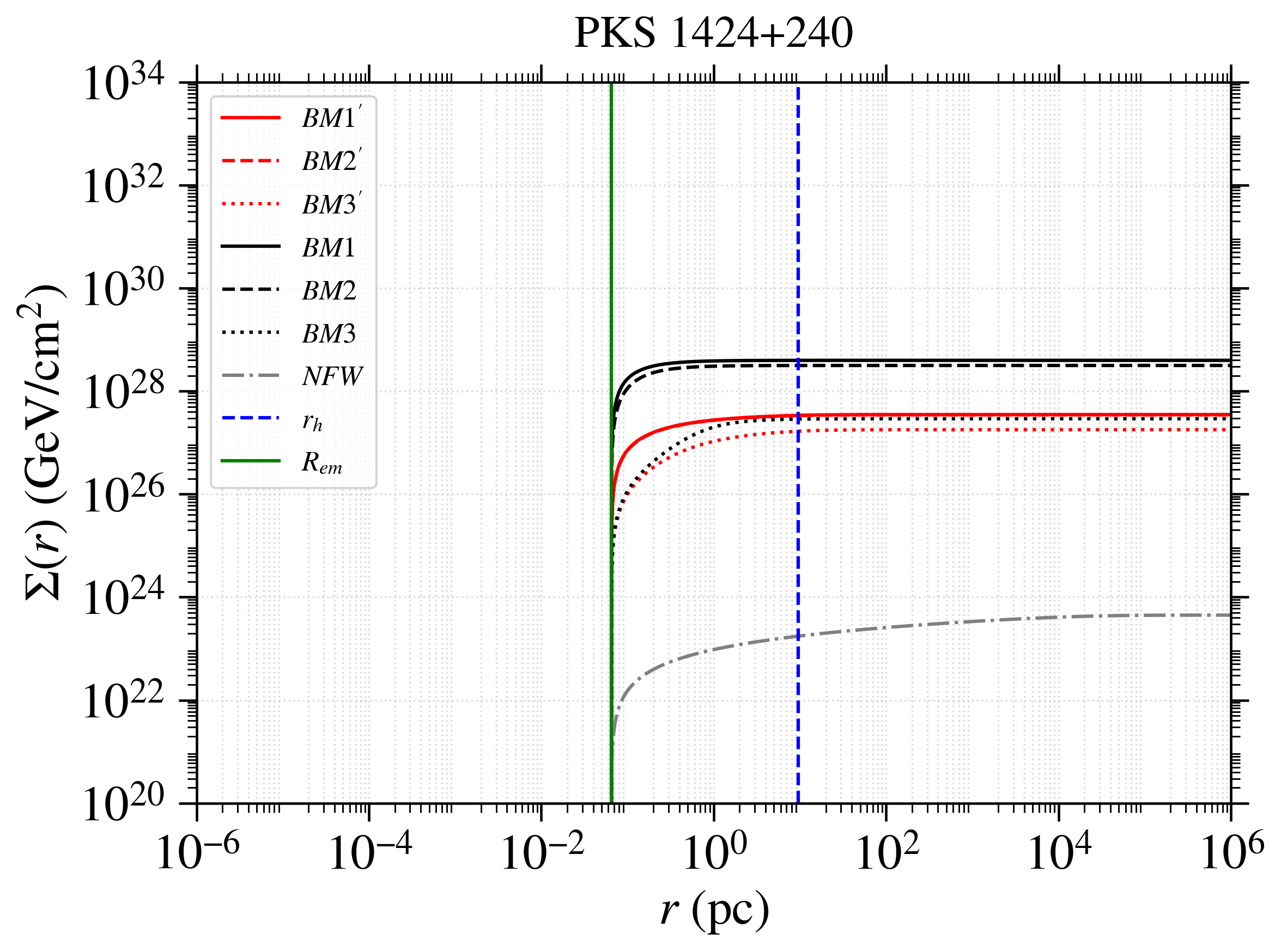}
\caption{DM column density $\Sigma$ Vs. $r$ (the upper limit to the integration given in Eq. (\ref{eq:Sigma_chi})) for each source. We considered the lower limit $r_l\equiv R_{em}$ to the same integration as given in Table \ref{tab:parameters} for each source.
} 
\label{fig:SigmaVsr}
\end{figure*}

Figure \ref{fig:SigmaVsr} here shows DM column density as a function of  radius from the SMBH for different profiles calculated using Eq.~(\ref{eq:Sigma_chi}). The lower limit of the integration, $r_l$ is set equal to the radius of emission $R_{em}$.

\section{Neutrino event distributions}
\label{sec:nu_events}

\setcounter{figure}{0}
\renewcommand{\thefigure}{\thesection\arabic{figure}}   

\begin{figure*}[htbp]
\includegraphics[width=0.75\linewidth]{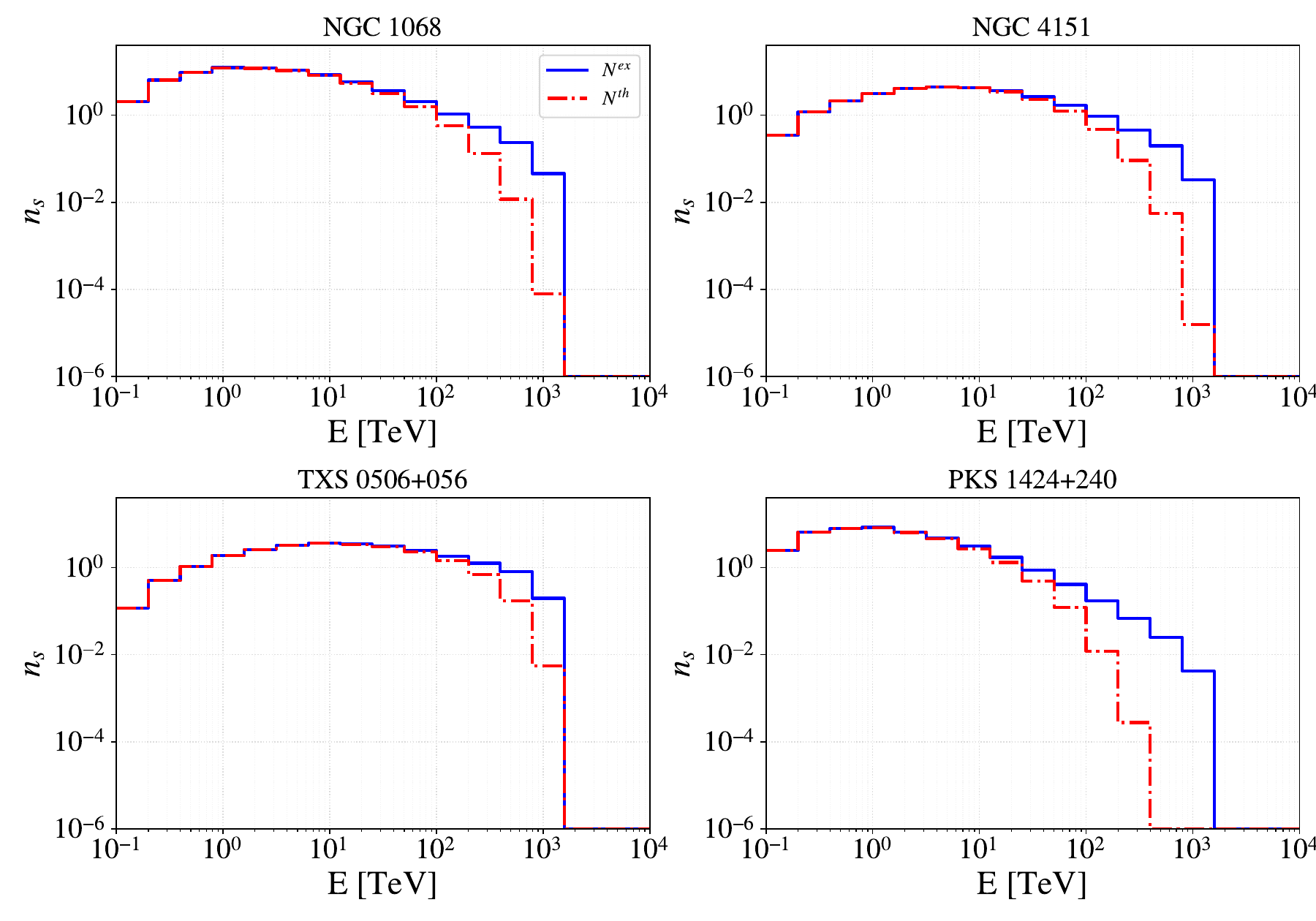}
\caption{Event distributions obtained from the IceCube data in the standard scenario (blue-solid) and in the DM-neutrino interaction scenario with energy-dependent cross-section at 90\% CL (red-dotdashed)}. \label{fig:eventdistribution}
\end{figure*}

 In Fig.~\ref{fig:eventdistribution} we show the event distributions in the energy-dependent cross-section case (see Eq.~(\ref{eq:discetizedcascade})) for the four sources we considered in our analysis. The values for the constant $A$ in Eq.~(\ref{eq:discetizedcascade}) for these plots are 0.0091 (NGC~1068), 0.0114 (NGC~4151), 0.0056 (TXS~0506+056) and 0.0311 (PKS~1424+240). We utilized the spectral index $\Gamma$ values from Table \ref{tab:n_s} to generate the $N_{ex}$ (blue-solid) distribution for each source. To obtain the $N_{th}$ (red-dashed) distribution, we profiled over $\Gamma$. Our findings revealed the following spectral index values: $\Gamma = 2.73$ for NGC 1068, $2.34$ for NGC 4151, $2.16$ for TXS 0506+056, and $2.96$ for PKS 1424+240. These values show only slight variations when compared to the observed $\Gamma$ values reported in Table \ref{tab:n_s}.

\section{Impact of uncertainty in $R_{em}$}
\label{sec:Rem_uncertainties}

\setcounter{figure}{0}
\renewcommand{\thefigure}{\thesection\arabic{figure}}   

\begin{figure*}[htbp]
\includegraphics[width=0.47\linewidth]{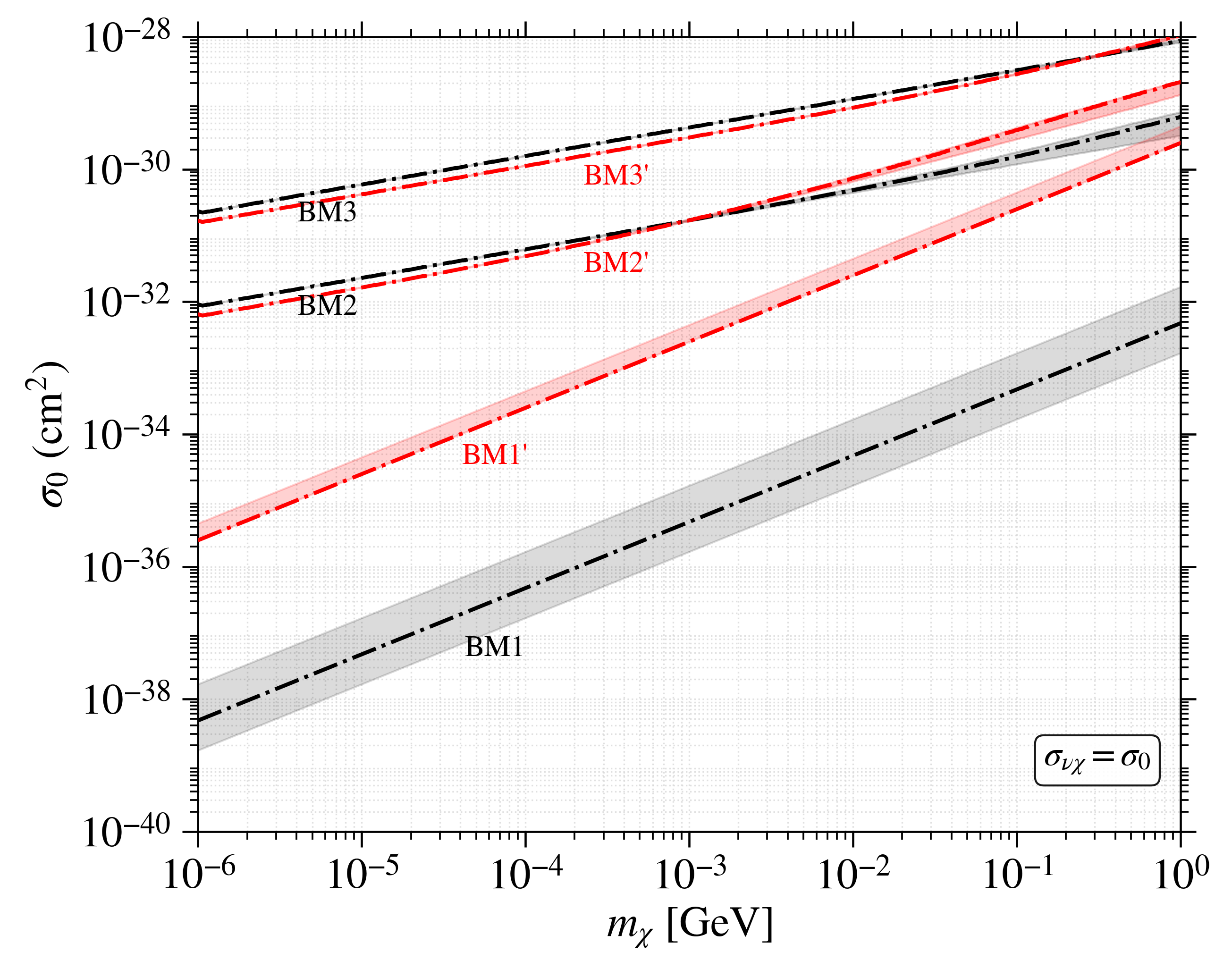}
\includegraphics[width=0.47\linewidth]{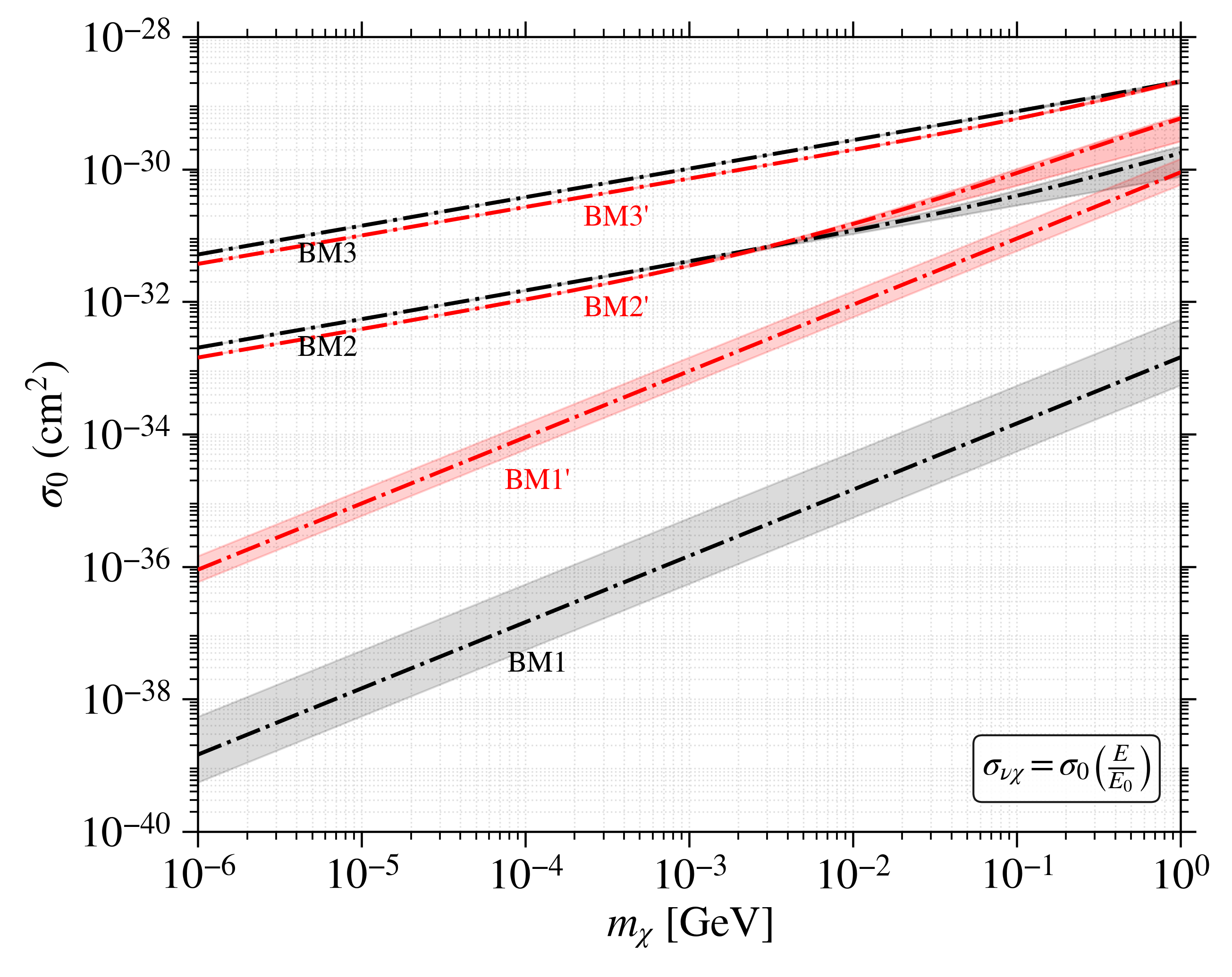}
\caption{The uncertainty band over $\sigma_0$ bounds vs.\ $m_\chi$ for stacking analysis covered for different DM density profiles, within the allowed range of $R_{em}$ for each source, for constant (left) and energy-dependent (right) $\sigma_{\nu\chi}$. The dot-dashed curves represent the stacking results obtained for fixed values of $R_{em}$ for each source reported in Table \ref{tab:parameters}. 
} 
\label{fig:UncertainRem}
\end{figure*}

Figure \ref{fig:UncertainRem} here shows uncertainty of the stacked constraints on $\sigma_0$ due to the uncertainty in neutrino emission radius (shaded regions in Fig.~\ref{fig:SigmaVsrl}) for individual AGNs analyzed here.

\section{Results of unbroken and broken power-law initial spectra on $\sigma_0$ constraints}
\label{sec:UP_BP_uncertainties}

\setcounter{figure}{0}
\renewcommand{\thefigure}{\thesection\arabic{figure}}   

\begin{figure*}
\centering
    \includegraphics[width=0.47\linewidth]{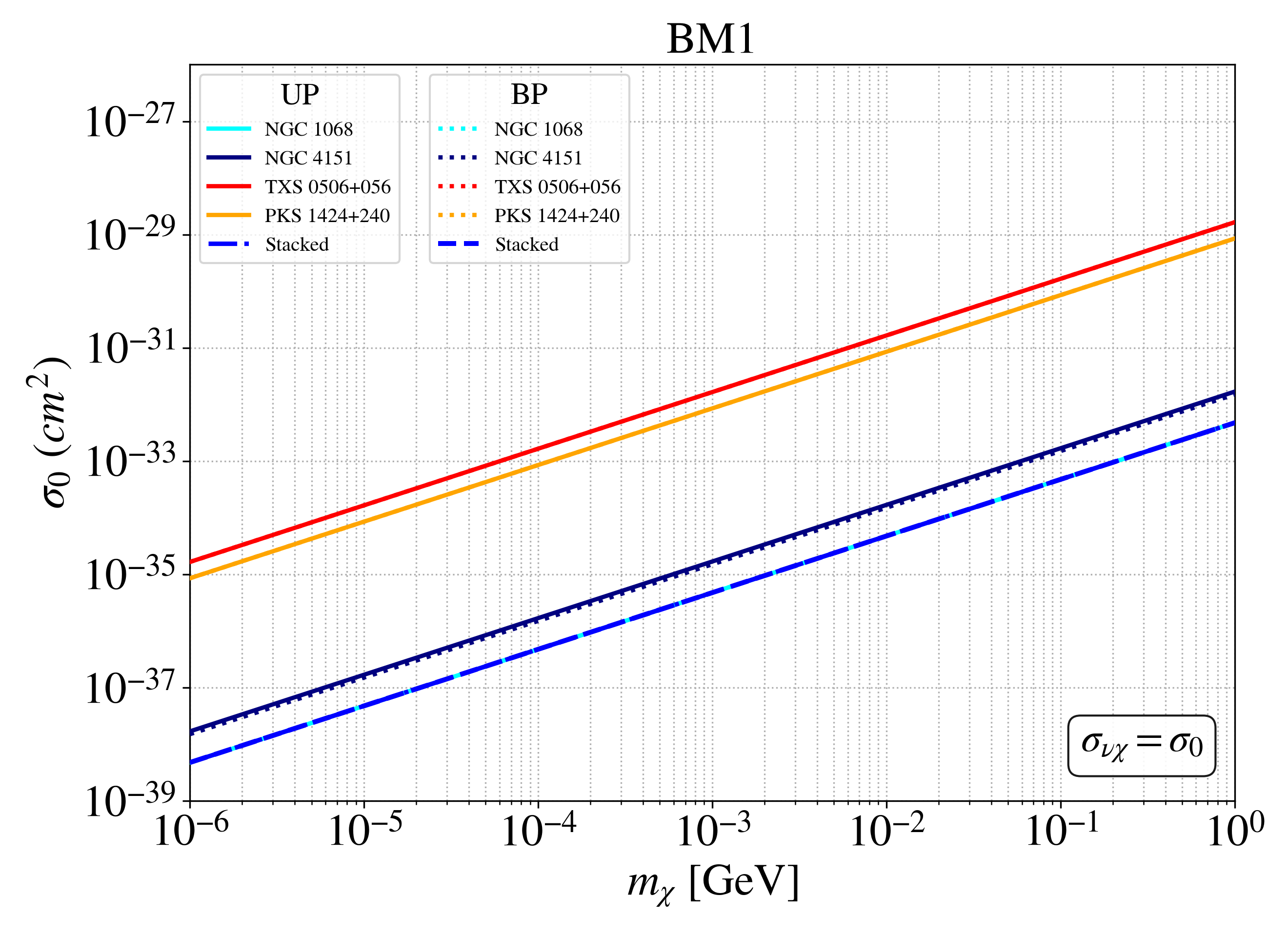}
    \includegraphics[width=0.47\linewidth]{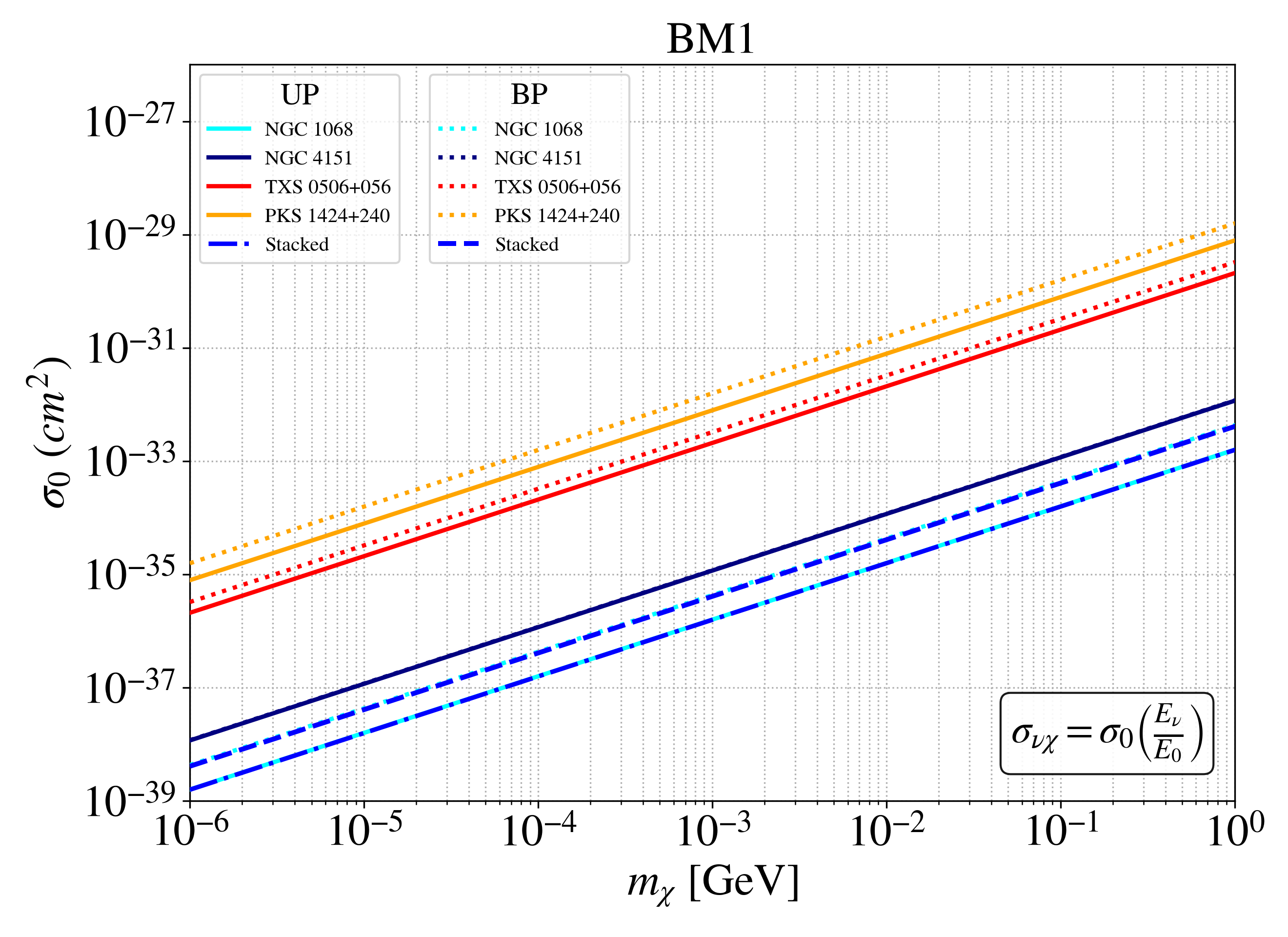}
    \caption{The solid and dotted curves correspond to constraints from individual sources in the case of UP and BP initial spectrum, respectively. The stacked results are represented by dot-dashed and dashed blue curves for the UP and BP, respectively. 
    }
    \label{fig:BrokenUnbroken}
\end{figure*}

The results of changing the initial neutrino spectrum from UP in Eq.~(\ref{eq:nuflux}) to BP in Eq.~(\ref{eq:nufluxbroken}) on the stacked $\sigma_0$ constraints are shown in
Fig.~\ref{fig:BrokenUnbroken}. We show results for the BM1 DM density profile only, since that gives the strongest constraint on $\sigma_0$.

\twocolumngrid

\nocite{*}


\begin{thebibliography}{10}

\bibitem{Bertone:2016nfn}
Gianfranco Bertone and Dan Hooper.
\newblock {History of dark matter}.
\newblock {\em Rev. Mod. Phys.}, 90(4):045002, 2018.

\bibitem{cirelli2024dark}
Marco Cirelli, Alessandro Strumia, and Jure Zupan.
\newblock Dark matter.
\newblock {\em arXiv preprint arXiv:2406.01705}, 2024.

\bibitem{Arguelles:2017atb}
Carlos~A. Arg{\"u}elles, Ali Kheirandish, and Aaron~C. Vincent.
\newblock {Imaging Galactic Dark Matter with High-Energy Cosmic Neutrinos}.
\newblock {\em Phys. Rev. Lett.}, 119(20):201801, 2017.

\bibitem{Choi:2019ixb}
Ki-Young Choi, Jongkuk Kim, and Carsten Rott.
\newblock {Constraining dark matter-neutrino interactions with IceCube-170922A}.
\newblock {\em Phys. Rev. D}, 99(8):083018, 2019.

\bibitem{Hooper:2021rjc}
Deanna~C. Hooper and Matteo Lucca.
\newblock {Hints of dark matter-neutrino interactions in Lyman-{\ensuremath{\alpha}} data}.
\newblock {\em Phys. Rev. D}, 105(10):103504, 2022.

\bibitem{PhysRevLett.130.091402}
James~M. Cline, Shan Gao, Fangyi Guo, Zhongan Lin, Shiyan Liu, Matteo Puel, Phillip Todd, and Tianzhuo Xiao.
\newblock Blazar constraints on neutrino-dark matter scattering.
\newblock {\em Phys. Rev. Lett.}, 130:091402, Feb 2023.

\bibitem{Cline_2023}
James~M. Cline and Matteo Puel.
\newblock Ngc 1068 constraints on neutrino-dark matter scattering.
\newblock {\em Journal of Cosmology and Astroparticle Physics}, 2023(06):004, jun 2023.

\bibitem{ferrer2023new}
Francesc Ferrer, Gonzalo Herrera, and Alejandro Ibarra.
\newblock New constraints on the dark matter-neutrino and dark matter-photon scattering cross sections from txs 0506+ 056.
\newblock {\em Journal of Cosmology and Astroparticle Physics}, 2023(05):057, 2023.

\bibitem{keivani2018multimessenger}
A~Keivani, K~Murase, M~Petropoulou, DB~Fox, SB~Cenko, S~Chaty, A~Coleiro, JJ~DeLaunay, S~Dimitrakoudis, PA~Evans, et~al.
\newblock A multimessenger picture of the flaring blazar txs 0506+ 056: implications for high-energy neutrino emission and cosmic-ray acceleration.
\newblock {\em The Astrophysical Journal}, 864(1):84, 2018.

\bibitem{Reimer:2018vvw}
Anita Reimer, Markus Boettcher, and Sara Buson.
\newblock {Cascading Constraints from Neutrino-emitting Blazars: The Case of TXS 0506+056}.
\newblock {\em Astrophys. J.}, 881(1):46, 2019.
\newblock [Erratum: Astrophys.J. 899, 168 (2020)].

\bibitem{Rodrigues:2018tku}
Xavier Rodrigues, Shan Gao, Anatoli Fedynitch, Andrea Palladino, and Walter Winter.
\newblock {Leptohadronic Blazar Models Applied to the 2014{\textendash}2015 Flare of TXS 0506+056}.
\newblock {\em Astrophys. J. Lett.}, 874(2):L29, 2019.

\bibitem{Cerruti:2018tmc}
M.~Cerruti, A.~Zech, C.~Boisson, G.~Emery, S.~Inoue, and J.~P. Lenain.
\newblock {Leptohadronic single-zone models for the electromagnetic and neutrino emission of TXS 0506+056}.
\newblock {\em Mon. Not. Roy. Astron. Soc.}, 483(1):L12--L16, 2019.
\newblock [Erratum: Mon.Not.Roy.Astron.Soc. 502, L21--L22 (2021)].

\bibitem{Inoue_2019}
Yoshiyuki Inoue, Dmitry Khangulyan, Susumu Inoue, and Akihiro Doi.
\newblock On high-energy particles in accretion disk coronae of supermassive black holes: Implications for mev gamma-rays and high-energy neutrinos from agn cores.
\newblock {\em The Astrophysical Journal}, 880(1):40, jul 2019.

\bibitem{Das:2021cdf}
Saikat Das, Soebur Razzaque, and Nayantara Gupta.
\newblock {Cosmogenic gamma-ray and neutrino fluxes from blazars associated with IceCube events}.
\newblock {\em Astron. Astrophys.}, 658:L6, 2022.

\bibitem{Das:2022nyp}
Saikat Das, Nayantara Gupta, and Soebur Razzaque.
\newblock {Implications of multiwavelength spectrum on cosmic-ray acceleration in blazar TXS 0506+056}.
\newblock {\em Astron. Astrophys.}, 668:A146, 2022.

\bibitem{IceCube:2021slf}
R.~Abbasi et~al.
\newblock {Search for Multi-flare Neutrino Emissions in 10 yr of IceCube Data from a Catalog of Sources}.
\newblock {\em Astrophys. J. Lett.}, 920(2):L45, 2021.


\bibitem{IceCube:2018dnn}
M.~G. Aartsen et~al.
\newblock {Multimessenger observations of a flaring blazar coincident with high-energy neutrino IceCube-170922A}.
\newblock {\em Science}, 361(6398):eaat1378, 2018.

\bibitem{IceCube:2018cha}
M.~G. Aartsen et~al.
\newblock {Neutrino emission from the direction of the blazar TXS 0506+056 prior to the IceCube-170922A alert}.
\newblock {\em Science}, 361(6398):147--151, 2018.

\bibitem{icecube2022evidence}
IceCube Collaboration*†, R~Abbasi, M~Ackermann, J~Adams, JA~Aguilar, M~Ahlers, M~Ahrens, JM~Alameddine, C~Alispach, AA~Alves~Jr, et~al.
\newblock Evidence for neutrino emission from the nearby active galaxy ngc 1068.
\newblock {\em Science}, 378(6619):538--543, 2022.

\bibitem{abbasi2025search}
R~Abbasi, M~Ackermann, J~Adams, SK~Agarwalla, JA~Aguilar, M~Ahlers, JM~Alameddine, NM~Amin, K~Andeen, C~Arg{\"u}elles, et~al.
\newblock Search for neutrino emission from hard X-ray AGN with IceCube.
\newblock {\em The Astrophysical Journal}, 981(2):131, 2025.

\bibitem{IceCube:2024ayt}
R.~Abbasi et~al.
\newblock {Search for Neutrino Emission from Hard X-Ray AGN with IceCube}.
\newblock {\em Astrophys. J.}, 981(2):131, 2025.


\bibitem{dixit2024searching}
K.~Dixit, L.~S.~Miranda and S.~Razzaque,
\newblock{Searching for pseudo-Dirac neutrinos from astrophysical sources in IceCube data}. 
{\em Eur. Phys. J. C} \textbf{85}, no.12, 1481 (2025).

\bibitem{Navarro:1995iw}
Julio~F. Navarro, Carlos~S. Frenk, and Simon D.~M. White.
\newblock {The Structure of cold dark matter halos}.
\newblock {\em Astrophys. J.}, 462:563--575, 1996.

\bibitem{Kravtsov:1997vm}
Andrey~V. Kravtsov, Anatoly~A. Klypin, and Alexei~M. Khokhlov.
\newblock {Adaptive refinement tree: A New high resolution N body code for cosmological simulations}.
\newblock {\em Astrophys. J. Suppl.}, 111:73, 1997.

\bibitem{Bryan:1997dn}
G.~L. Bryan and M.~L. Norman.
\newblock {Statistical properties of x-ray clusters: Analytic and numerical comparisons}.
\newblock {\em Astrophys. J.}, 495:80, 1998.

\bibitem{WMAP:2006bqn}
D.~N. Spergel et~al.
\newblock {Wilkinson Microwave Anisotropy Probe (WMAP) three year results: implications for cosmology}.
\newblock {\em Astrophys. J. Suppl.}, 170:377, 2007.

\bibitem{Gondolo:1999ef}
Paolo Gondolo and Joseph Silk.
\newblock {Dark matter annihilation at the galactic center}.
\newblock {\em Phys. Rev. Lett.}, 83:1719--1722, 1999.

\bibitem{Zhao:1995cp}
HongSheng Zhao.
\newblock {Analytical models for galactic nuclei}.
\newblock {\em Mon. Not. Roy. Astron. Soc.}, 278:488--496, 1996.

\bibitem{Stahl:2024jzk}
Cl{\'e}ment Stahl, Nicolas Mai, Benoit Famaey, Yohan Dubois, and Rodrigo Ibata.
\newblock {From inflation to dark matter halo profiles: the impact of primordial non-Gaussianities on the central density~cusp}.
\newblock {\em JCAP}, 05:021, 2024.

\bibitem{PhysRevLett.92.201304}
David Merritt.
\newblock Evolution of the dark matter distribution at the galactic center.
\newblock {\em Phys. Rev. Lett.}, 92:201304, May 2004.

\bibitem{PhysRevLett.93.061302}
Oleg~Y. Gnedin and Joel~R. Primack.
\newblock Dark matter profile in the galactic center.
\newblock {\em Phys. Rev. Lett.}, 93:061302, Aug 2004.

\bibitem{PhysRevD.75.043517}
David Merritt, Stefan Harfst, and Gianfranco Bertone.
\newblock Collisionally regenerated dark matter structures in galactic nuclei.
\newblock {\em Phys. Rev. D}, 75:043517, Feb 2007.

\bibitem{PhysRevD.106.043018}
Stuart~L. Shapiro and Douglas~C. Heggie.
\newblock Effect of stars on the dark matter spike around a black hole: A tale of two treatments.
\newblock {\em Phys. Rev. D}, 106:043018, Aug 2022.

\bibitem{Vasiliev:2008uz}
Eugene Vasiliev and Maxim Zelnikov.
\newblock {Dark matter dynamics in Galactic center}.
\newblock {\em Phys. Rev. D}, 78:083506, 2008.

\bibitem{Greenhill:1996te}
L.~J. Greenhill, C.~R. Gwinn, R.~Antonucci, and R.~Barvainis.
\newblock {Vlbi imaging of water maser emission from the nuclear torus of ngc 1068}.
\newblock {\em Astrophys. J. Lett.}, 472:L21, 1996.

\bibitem{Bentz_2022}
Misty~C. Bentz, Peter~R. Williams, and Tommaso Treu.
\newblock The broad line region and black hole mass of ngc 4151.
\newblock {\em The Astrophysical Journal}, 934(2):168, aug 2022.

\bibitem{Padovani:2019xcv}
P.~Padovani, F.~Oikonomou, M.~Petropoulou, P.~Giommi, and E.~Resconi.
\newblock {TXS 0506+056, the first cosmic neutrino source, is not a BL Lac}.
\newblock {\em Mon. Not. Roy. Astron. Soc.}, 484(1):L104--L108, 2019.

\bibitem{Cerruti:2017mnz}
M.~Cerruti, W.~Benbow, X.~Chen, J.~P. Dumm, L.~F. Fortson, and K.~Shahinyan.
\newblock {Luminous and high-frequency peaked blazars: the origin of the {\ensuremath{\gamma}}-ray emission from PKS 1424+240}.
\newblock {\em Astron. Astrophys.}, 606:A68, 2017.

\bibitem{granelli2022blazar}
Alessandro Granelli, Piero Ullio, and Jin-Wei Wang.
\newblock Blazar-boosted dark matter at super-kamiokande.
\newblock {\em Journal of Cosmology and Astroparticle Physics}, 2022(07):013, 2022.

\bibitem{Murase:2019vdl}
Kohta Murase, Shigeo~S. Kimura, and Peter Meszaros.
\newblock {Hidden Cores of Active Galactic Nuclei as the Origin of Medium-Energy Neutrinos: Critical Tests with the MeV Gamma-Ray Connection}.
\newblock {\em Phys. Rev. Lett.}, 125(1):011101, 2020.

\bibitem{Inoue:2019yfs}
Yoshiyuki Inoue, Dmitry Khangulyan, and Akihiro Doi.
\newblock {On the Origin of High-energy Neutrinos from NGC 1068: The Role of Nonthermal Coronal Activity}.
\newblock {\em Astrophys. J. Lett.}, 891(2):L33, 2020.

\bibitem{Inoue:2018kbv}
Yoshiyuki Inoue and Akihiro Doi.
\newblock {Detection of Coronal Magnetic Activity in Nearby Active Supermassive Black Holes}.
\newblock {\em Astrophys. J.}, 869(2):114, 2018.

\bibitem{Gallimore:2004wk}
Jack~F. Gallimore, Stefi~A. Baum, and Christopher~P. O'Dea.
\newblock {The parsec-scale radio structure of ngc 1068 and the nature of the nuclear radio source}.
\newblock {\em Astrophys. J.}, 613:794--810, 2004.

\bibitem{Murase:2022dog}
Kohta Murase.
\newblock {Hidden Hearts of Neutrino Active Galaxies}.
\newblock {\em Astrophys. J. Lett.}, 941(1):L17, 2022.

\bibitem{10.1093/mnras/staa3363}
Olmo Piana, Pratika Dayal, Marta Volonteri, and Tirthankar~Roy Choudhury.
\newblock The mass assembly of high-redshift black holes.
\newblock {\em Monthly Notices of the Royal Astronomical Society}, 500(2):2146--2158, 10 2020.

\bibitem{Hure:2002nu}
Jean-Marc Hure.
\newblock {Origin of non-keplerian motions of masers in ngc 1068}.
\newblock {\em Astron. Astrophys.}, 395:L21--L24, 2002.

\bibitem{Lodato:2002cv}
Giuseppe Lodato and G.~Bertin.
\newblock {Non-Keplerian rotation in the nucleus of NGC 1068: Evidence for a massive accretion disk?}
\newblock {\em Astron. Astrophys.}, 398:517--524, 2003.

\bibitem{Woo:2002un}
Jong-Hak Woo and C.~Megan Urry.
\newblock {AGN black hole masses and bolometric luminosities}.
\newblock {\em Astrophys. J.}, 579:530--544, 2002.

\bibitem{Panessa:2006sg}
Francesca Panessa, L.~Bassani, M.~Cappi, M.~Dadina, X.~Barcons, F.~J. Carrera, L.~C. Ho, and K.~Iwasawa.
\newblock {On the X-ray, optical emission line and black hole mass properties of local Seyfert galaxies}.
\newblock {\em Astron. Astrophys.}, 455:173, 2006.

\bibitem{2014ApJ...791...37O}
Christopher~A. {Onken}, Monica {Valluri}, Jonathan~S. {Brown}, Peter~J. {McGregor}, Bradley~M. {Peterson}, Misty~C. {Bentz}, Laura {Ferrarese}, Richard~W. {Pogge}, Marianne {Vestergaard}, Thaisa {Storchi-Bergmann}, and Rogemar~A. {Riffel}.
\newblock {The Black Hole Mass of NGC 4151. II. Stellar Dynamical Measurement from Near-infrared Integral Field Spectroscopy}.
\newblock {\em \apj}, 791(1):37, August 2014.

\bibitem{PhysRevD.82.083514}
Mikhail Gorchtein, Stefano Profumo, and Lorenzo Ubaldi.
\newblock Probing dark matter with active galactic nuclei jets.
\newblock {\em Phys. Rev. D}, 82:083514, Oct 2010.

\bibitem{Sadeghian:2013laa}
Laleh Sadeghian, Francesc Ferrer, and Clifford~M. Will.
\newblock {Dark matter distributions around massive black holes: A general relativistic analysis}.
\newblock {\em Phys. Rev. D}, 88(6):063522, 2013.

\bibitem{DiMatteo:2003zx}
Tiziana Di~Matteo, Rupert A.~C. Croft, Volker Springel, and Lars Hernquist.
\newblock {Black hole growth and activity in a lambda CDM universe}.
\newblock {\em Astrophys. J.}, 593:56--68, 2003.

\bibitem{Ferrarese:2002ct}
Laura Ferrarese.
\newblock {Beyond the bulge: a fundamental relation between supermassive black holes and dark matter halos}.
\newblock {\em Astrophys. J.}, 578:90--97, 2002.

\bibitem{Baes:2003rt}
Maarten Baes, Pieter Buyle, George K.~T. Hau, and Herwig Dejonghe.
\newblock {Observational evidence for a connection between supermassive black holes and dark matter haloes}.
\newblock {\em Mon. Not. Roy. Astron. Soc.}, 341:L44, 2003.

\bibitem{Vincent:2017svp}
Aaron~C. Vincent, Carlos~A. Arg{\"u}elles, and Ali Kheirandish.
\newblock {High-energy neutrino attenuation in the Earth and its associated uncertainties}.
\newblock {\em JCAP}, 11:012, 2017.

\bibitem{Lin:2022dbl}
Yen-Hsun Lin, Wen-Hua Wu, Meng-Ru Wu, and Henry Tsz-King Wong.
\newblock {Searching for Afterglow: Light Dark Matter Boosted by Supernova Neutrinos}.
\newblock {\em Phys. Rev. Lett.}, 130(11):111002, 2023.

\bibitem{Jho:2021rmn}
Yongsoo Jho, Jong-Chul Park, Seong~Chan Park, and Po-Yan Tseng.
\newblock {Cosmic-Neutrino-Boosted Dark Matter ($\nu$BDM)}.
\newblock 1 2021.

\bibitem{Ghosh:2021vkt}
Diptimoy Ghosh, Atanu Guha, and Divya Sachdeva.
\newblock {Exclusion limits on dark matter-neutrino scattering cross section}.
\newblock {\em Phys. Rev. D}, 105(10):103029, 2022.

\bibitem{Mosbech:2020ahp}
Markus~R. Mosbech, Celine Boehm, Steen Hannestad, Olga Mena, Julia Stadler, and Yvonne Y.~Y. Wong.
\newblock {The full Boltzmann hierarchy for dark matter-massive neutrino interactions}.
\newblock {\em JCAP}, 03:066, 2021.

\bibitem{baker1984clarification}
Steve Baker and Robert~D Cousins.
\newblock Clarification of the use of chi-square and likelihood functions in fits to histograms.
\newblock {\em Nuclear Instruments and Methods in Physics Research}, 221(2):437--442, 1984.

\bibitem{Gasparyan:2021oad}
Sargis Gasparyan, Damien B{\'e}gu{\'e}, and Narek Sahakyan.
\newblock {Time-dependent lepto-hadronic modelling of the emission from blazar jets with SOPRANO: the case of TXS 0506~+~056, 3HSP J095507.9~+~355101, and 3C 279}.
\newblock {\em Mon. Not. Roy. Astron. Soc.}, 509(2):2102--2121, 2021.

\bibitem{He:1990pn}
X.~G. He, Girish~C. Joshi, H.~Lew, and R.~R. Volkas.
\newblock {NEW Z-prime PHENOMENOLOGY}.
\newblock {\em Phys. Rev. D}, 43:22--24, 1991.

\bibitem{He:1991qd}
Xiao-Gang He, Girish~C. Joshi, H.~Lew, and R.~R. Volkas.
\newblock {Simplest Z-prime model}.
\newblock {\em Phys. Rev. D}, 44:2118--2132, 1991.

\bibitem{Araki:2017wyg}
Takeshi Araki, Shihori Hoshino, Toshihiko Ota, Joe Sato, and Takashi Shimomura.
\newblock {Detecting the $L_{\mu}-L_{\tau}$ gauge boson at Belle II}.
\newblock {\em Phys. Rev. D}, 95(5):055006, 2017.

\bibitem{Bernal:2025szh}
Nicol{\'a}s Bernal, Jacinto~P. Neto, Javier Silva-Malpartida, and Farinaldo~S. Queiroz.
\newblock {Enabling thermal dark matter within the vanilla L{\ensuremath{\mu}}-L{\ensuremath{\tau}} model}.
\newblock {\em Phys. Rev. D}, 112(7):075042, 2025.

\bibitem{Kahn:2018cqs}
Yonatan Kahn, Gordan Krnjaic, Nhan Tran, and Andrew Whitbeck.
\newblock {M$^{3}$: a new muon missing momentum experiment to probe (g {\ensuremath{-}} 2)$_{\mu}$ and dark matter at Fermilab}.
\newblock {\em JHEP}, 09:153, 2018.

\bibitem{Buckley:2022btu}
Matthew~R. Buckley, Andrew Mastbaum, and Gopolang Mohlabeng.
\newblock {Directional neutrino searches for Galactic Center dark matter at large underground LArTPCs}.
\newblock {\em Phys. Rev. D}, 107(9):092006, 2023.

\bibitem{Planck:2018vyg}
N.~Aghanim et~al.
\newblock {Planck 2018 results. VI. Cosmological parameters}.
\newblock {\em Astron. Astrophys.}, 641:A6, 2020.
\newblock [Erratum: Astron.Astrophys. 652, C4 (2021)].

\bibitem{Wells:1994qy}
James~D. Wells.
\newblock {Annihilation cross-sections for relic densities in the low velocity limit}.
\newblock 3 1994.

\bibitem{Steigman:2012nb}
Gary Steigman, Basudeb Dasgupta, and John~F. Beacom.
\newblock {Precise Relic WIMP Abundance and its Impact on Searches for Dark Matter Annihilation}.
\newblock {\em Phys. Rev. D}, 86:023506, 2012.

\bibitem{Murase:2019xqi}
Kohta Murase and Ian~M. Shoemaker.
\newblock {Neutrino Echoes from Multimessenger Transient Sources}.
\newblock {\em Phys. Rev. Lett.}, 123(24):241102, 2019.

\bibitem{Koren:2019wwi}
Seth Koren.
\newblock {Neutrino -- Dark Matter Scattering and Coincident Detections of UHE Neutrinos with EM Sources}.
\newblock {\em JCAP}, 09:013, 2019.

\bibitem{McMullen:2021ikf}
Adam McMullen, Aaron Vincent, Carlos Arguelles, and Austin Schneider.
\newblock {Dark matter neutrino scattering in the galactic centre with IceCube}.
\newblock {\em JINST}, 16(08):C08001, 2021.

\bibitem{Fayet:2006sa}
Pierre Fayet, Dan Hooper, and Gunter Sigl.
\newblock {Constraints on light dark matter from core-collapse supernovae}.
\newblock {\em Phys. Rev. Lett.}, 96:211302, 2006.

\bibitem{mangano2006cosmological}
Gianpiero Mangano, Alessandro Melchiorri, Paolo Serra, Asantha Cooray, and Marc Kamionkowski.
\newblock Cosmological bounds on dark-matter-neutrino interactions.
\newblock {\em Physical Review D—Particles, Fields, Gravitation, and Cosmology}, 74(4):043517, 2006.

\bibitem{boehm2013lower}
C{\'e}line Boehm, Matthew~J Dolan, and Christopher McCabe.
\newblock A lower bound on the mass of cold thermal dark matter from planck.
\newblock {\em Journal of Cosmology and Astroparticle Physics}, 2013(08):041, 2013.

\bibitem{2014arXiv1412.3113B}
Bridget {Bertoni}, Seyda {Ipek}, David {McKeen}, and Ann~E. {Nelson}.
\newblock {Reducing cosmological small scale structure via a large dark matter-neutrino interaction: constraints and consequences}.
\newblock {\em arXiv e-prints}, page arXiv:1412.3113, December 2014.

\bibitem{wilkinson2014constraining}
Ryan~J Wilkinson, Celine Boehm, and Julien Lesgourgues.
\newblock Constraining dark matter-neutrino interactions using the cmb and large-scale structure.
\newblock {\em Journal of Cosmology and Astroparticle Physics}, 2014(05):011, 2014.

\bibitem{Gentile:2007sb}
Gianfranco Gentile, Chiara Tonini, and Paolo Salucci.
\newblock {Lambda CDM Halo Density Profiles: Where do actual halos converge to NFW ones?}
\newblock {\em Astron. Astrophys.}, 467:925--931, 2007.


\bibitem{Akita:2023yga}
Kensuke Akita1, Shin’ichiro Ando,
\newblock {Constraints on dark matter-neutrino scattering from the Milky-Way satellites and subhalo modeling for dark acoustic oscillations},
\newblock {\em Journal of Cosmology and Astroparticle Physics}, \textbf{11}, 037 (2023).

\bibitem{Bertolez-Martinez:2025trs}
T.~Bert{\'o}lez-Mart{\'\i}nez, G.~Herrera, P.~Mart{\'\i}nez-Mirav{\'e} and J.~Terol Calvo,
\newblock {The Highest-Energy Neutrino Event Constrains Dark Matter-Neutrino Interactions},
\newblock {\em arXiv e-prints}, page arXiv:2506.08993, June 2025.
[arXiv:2506.08993 [hep-ph]].

\bibitem{Mondol:2025uuw}
R.~Mondol, S.~Bouri, A.~K.~Saha and R.~Laha,
[arXiv:2506.19910 [hep-ph]].

\bibitem{Dev:2025tdv}
P.~S.~B.~Dev, D.~Kim, D.~Sathyan, K.~Sinha and Y.~Zhang,
\newblock {New Constraints on Neutrino-Dark Matter Interactions: A Comprehensive Analysis},
[arXiv:2507.01000 [hep-ph]].

\end{thebibliography}

\end{document}